\documentclass[manuscript]{rmaa} \usepackage{paralist}
\usepackage{psfrag,color} \usepackage[latin1]{inputenc}
\usepackage{graphics}

 \newcommand{\hii}{\ion{H}{2}}

	\title{ Chemical Evolution and the Galactic Habitable Zone of M31
(the Andromeda Galaxy)
 } 
	\author{ Leticia Carigi,\altaffilmark{1} Jorge Garc\'{\i}a-Rojas,
\altaffilmark{2,3} and Sof\'{\i}a Meneses-Goytia \altaffilmark{1, 4}
}
	\altaffiltext{1}{ Instituto de Astronom\'{\i}a, Universidad
Nacional Aut\'onoma de M\'exico. Apartado Postal 70-264, Ciudad
Universitaria, M\'exico DF 04510, M\'exico.
}
	\altaffiltext{2}{ Instituto de Astrof\'{\i}sica de Canarias.
E-38200. La Laguna, Tenerife, Spain.
}
	\altaffiltext{3}{ Departamento de Astrof\'{\i}sica, Universidad
de La Laguna. E-38205. La Laguna, Spain.
}
	  \altaffiltext{4}{ Kapteyn Instituut, Rijkuniversiteit
  Groningen. Landleven 12, 9747 AD, Groningen, Nederland.
}
	\shortauthor{Carigi, Garc\'{\i}a-Rojas, \& Meneses-Goytia}
\shorttitle{Chemical Evolution and GHZ of M31} \fulladdresses{ \item
Leticia Carigi: Instituto de Astronom\'{\i}a, Universidad Nacional
Aut\'onoma de M\'exico. Apartado Postal 70-264, Ciudad Universitaria,
M\'exico DF 04510, M\'exico (carigi@astro.unam.mx). \item Jorge
Garc\'{\i}a-Rojas: Instituto de Astrof\'{\i}sica de Canarias.
E-38200. La Laguna, Tenerife, Spain. Departamento de
Astrof\'{\i}sica, Universidad de La Laguna. E-38205. La Laguna, Spain
(jogarcia@iac.es). \item Sof\'{\i}a Meneses-Goytia: Kapteyn
Instituut, Rijkuniversiteit Groningen. Landleven 12, 9747 AD,
Groningen, Nederland (s.meneses-goytia@astro.rug.nl).
}
	\listofauthors{L. Carigi, J. Garc\'{\i}a-Rojas, \& S.
Meneses-Goytia} \indexauthor{Carigi, L.}
\indexauthor{Garc\'{\i}a-Rojas, J.} \indexauthor{Meneses-Goytia, S.}

	\abstract { We have computed the Galactic Habitable Zones
(GHZs) of the Andromeda galaxy (M31) based on the probability of
terrestrial planet formation, which depends on the metallicity ($Z$)
of the interstellar medium, and the number of 
stars formed per unit surface area.
The GHZ was therefore obtained from a chemical
evolution model (CEM) built to reproduce a metallicity gradient in
the galactic disk, [O/H]($r$) $ = -0.015 \pm 0.003$ dex kpc$^{-1}
\times r$(kpc) $ + 0.44 \pm 0.04$ dex. This gradient is the most
probable when intrinsic scatter is present in the observational data.
The chemical evolution model predicted a higher star formation
history (SFH) in both the halo and disk components of M31 and a less
efficient inside-out galactic formation, compared to those of the
Milky Way. If we assumed that Earth-like planets form with a
probability law that follows the $Z$ distribution shown by stars with
detected planets and the SFH predicted by the CEM, the most
probable GHZ per $pc^2$ is located between  3 and 7 kpc for
planets with ages between 6 and 7 Gy, approximately. But the highest
number of stars with habitable planets is in a ring located between
12 and 14 kpc with mean age of $\sim$ 7 Gy. 11 \% and 6.5 \% of the
all formed stars in M31 may have planets capable of hosting
basic and complex life, respectively.
}
	\resumen { Calculamos la Zona de Habitabilidad Gal\'actica (ZHG)
de M31 bas\'andonos en la probabilidad de formaci\'on de planetas
terrestres dependiente de la metalicidad ($Z$) del medio
interestelar. La ZHG fue determinada a partir de un modelo de
evoluci\'on qu\'{\i}mica construido para reproducir un gradiente de
metalicidad  en el disco gal\'actico: [O/H]($r$)$ =-0.015$ dex
kpc$^{-1} \times r$(kpc)$ + 0.44$ dex. Suponiendo que los planetas
tipo Tierra se forman bajo una ley de probabilidad que sigue la
distribuci\'on de $Z$ mostrada por las estrellas con planetas,
entonces la ZHG m\'as probable se localiza entre 3 y 7 kpc para 
planetas con edades entre 6 y 7 Ga, aproximadamente, aunque el mayor
n\'umero de estrellas con planetas habitables se encuentra en un anillo
localizado entre 12 y 14 kpc y edad promedio de $\sim$ 7 Ga. El 11
\% y 6.5 \% de todas las estrellas formadas en M31 podr\'{\i}an albergar
vida b\'asica y compleja, respectivamente.
}
	\addkeyword{Galactic habitability} \addkeyword{Chemical
evolution} \addkeyword{Abundance gradients} \addkeyword{Andromeda
galaxy} \addkeyword{M31}

\begin{document} \maketitle

	\section{Introduction} \label{sec:intro}

	The Galactic Habitable Zone (GHZ) is defined as the region with
sufficient abundance of chemical elements to form planetary systems
in which Earth-like planets could be found and might be capable of
sustaining life (Gonzalez et al. 2001, Lineweaver 2001). An
Earth-like planet is a rocky planet characterized in general terms by
the presence of water and an atmosphere (Segura \& Kaltenegger 2009).

	GHZ research has focused mainly on our galaxy, the Milky Way (MW)
(Gonzalez et al. 2001, Lineweaver et al. 2004, Prantzos 2008,
Gowanlock et al. 2011). Gonzalez et al. were the first to propose the
concept of a GHZ, which is a ring located in the
thin disk that migrates outwards with time, due to their most
important assumption is that terrestrial planet form with
metallicities higher than 1/2 of the solar value ($Z_\odot$).
Lineweaver et al. (2004) later proposed that the Milky Way's GHZ is a
ring located in the Galactic disk within a radius interval of 7 to 9
kpc from the center of the MW, and that the area of the ring
increases with the age of the Galaxy, because they consider a
$Z$ distribution, for forming terrestrial planets, that peaks at  
$\sim 0.8 Z_\odot$; the presence of a host star; and the absence of
nearby supernovae (SN) harmful to life. On the other hand, Prantzos
concluded that the current GHZ covers practically the entire MW
disk, due to he assumes a $Z$ probability, to form Earth-like
planets, almost equal for $Z > 0.1 Z_\odot$. These studies confirmed
that the Solar System is located within the GHZ since the Sun is
found at 8 kpc from the center of the Galaxy. Nevertheless, a star
with an Earth-like planet capable of sustaining life is more likely
to be found in inner rings of the Galactic disk, between 2 and 4 kpc,
owing to the high stellar surface density that is present in the
inner disk of our galaxy (Prantzos 2008, Gowanlock et al. 2011).

	The GHZ has recently been computed for two elliptical galaxies
(Suthar \& McKay 2012). Imposing only metallicity restrictions for
planet formation, these authors found that both elliptical galaxies
could sustain broad GHZs.

	Here, we extend MW studies to the disk of the most massive galaxy
in the Local Group: the Andromeda galaxy (M31). M31 is a type SAb
spiral galaxy, whose visible mass is $\sim$1.2 times larger than
that of the MW, and M31 is at a distance of 783 $\pm$ 30 kpc
(Holland 1998).

	The GHZ depends mainly on the abundance of chemical elements
heavier than He (metallicity, $Z$), since $Z$ leads to planetary
formation. Moreover,  the GHZ also depends on the occurrence of 
strong radiation events that can sterilize a planet. Melott \& Thomas
(2011) study many kinds of astrophysical radiations lethal to life,
as electromagnetic radiation (e.g. X rays), high-energy protons or
cosmic rays from stars (included the Sun), SN, and
gamma-ray bursts (GRB). According to them, the SN and GRB could be the
dominant cause of extinctions. Since most of the SN and GRB
are originated during the last stages of massive stars, the rate of
high-mass stars is useful to estimate the galactic zones where life
on planets is annihilated by astrophysical events.
In this paper, we excluded the bulge of Andromeda from the GHZ,
despite its having a high enough abundance of chemical elements to
form planets (Sarajedini \& Jablonka 2005). The bulge, located
between 0 and 3 kpc from de galactic center, might not provide a
stable habitat for life, due to its high supernovae rate at early
times, which could sterilize planets, and the proximity of stars
that can destabilize the orbits of the planets (Jim\'enez-Torres et
al. 2011).

	Throughout this paper the terms "evolved life" and "complex life"
are synonymous. These terms refer to a type of life similar to that
of the human beings, since they are able to develop advanced
technologies.

	In this study, we present a chemical evolution model (CEM) for
the halo and disk components that predicts the temporal behavior of
the space distribution of $Z$ and SN occurrence in M31, built
on precise observational constraints (Sec. 2). Based on a set of
biogenic, astrophysical, and geophysical restrictions, the CEM
results lead to the determination of the GHZ (Sec. 3). We discuss the
implication of the chemical evolution model and the GHZ condition on
the location, size, and age of the Galactic Habitable Zone (Sec. 4).
Finally, we present our conclusions of the present study (Sec. 5).

	\section{CHEMICAL EVOLUTION MODEL, CEM} \label{sec:cem}

	Chemical evolution models study the changes, in space and time,
of: a) the chemical abundances present in the interstellar medium
(ISM), b) the gas mass, and c) the total baryonic mass of galaxies
and intergalactic medium. These studies have a considerable number of
free parameters and observational constraints allow us to estimate
some of them. If the number of observational constraints is high and
the observational data are precise, a more solid model could be
proposed and therefore a better estimate of the GHZ can be obtained.
In consequence, before building the CEM, we collected and determined
a reliable data set of observational constraints.

%
	\subsection{Observational constraints} \label{sec:observations}

	The present CEM was built to reproduce M31's three main
observational constraints of the galactic disk: the radial
distributions of the total baryonic mass, the gas mass, and the
oxygen abundance. Since the data come from several authors, who have
adopted different distances, the data used in this work were
corrected according to our adopted distance of 783 kpc for M31 \citep{holland98}.

%
	\subsubsection{Radial distribution of the total mass surface
density in the disk, $\Sigma_{T}(r)$} \label{sec:mtot}

	The luminosity profile is produced by the stellar and ionized
gaseous components of the galaxies. In evolved spiral galaxies, such
as M31, the current luminosity is mostly due to the stellar
component. Because the luminosity profiles of such galaxies follow an
exponential behavior with respect to galactocentric distance ($r$),
we associated that profile to the radial distribution of total mass
surface density, $\Sigma_{T}(r)=\Sigma_0 exp(-r/r_d)$ where $r_d =
5.5$ kpc (Renda et al. 2005). The value for $\Sigma_0$ is 548.2
M$_\odot pc^{-2}$ and was obtained by the integration of
$\Sigma_{T}(r)$ over the disk's surface in order to reproduce the
total mass disk  of M31, which is $7.2 \times  10^{10}$ M$_\odot$
(Widrow, Perrett \& Suyu 2003).

%
	\subsubsection{Radial distribution of oxygen abundance,
[O/H]($r$)} \label{sec:oh}

	The data were taken from the compilation of M31's ionized
hydrogen (HII) region spectra by Blair et al. (1982) and Galarza et
al. (1999). From the total region sample, we rejected those with
uncertain measurements of [OII], [OIII], and [NII]
lines.\footnote{The notation [XI] implies the atom X in neutral form,
[XII] indicates X$^+$, [XIII] is X$^{2+}$, etc} Taking into account
these prescriptions, the final number of HII regions of our sample is
equal to 83. Once we selected our sample, we employed the R$_{23}$
bi-evaluated method from Pagel et al. (1979), which links the
intensity of strong [OII] and [OIII] emission lines with [O/H], to
compute the abundances.

	The disadvantage of this method is that it is double-valued with
respect to metallicity. In fact, at low oxygen abundances ([O/H] $\le
-0.66$) the R$_{23}$ index decreases with the abundance, while for
high oxygen abundances ([O/H] $\ge -0.41$) the metals cooling
efficiency makes R$_{23}$ drop with rising abundance. In order to
break the R$_{23}$ method's degeneracy, we used [NII]/[OII] line
ratios, which are not sensitive to the ionization parameter and are a
strong function of O/H above log([NII]/[OII]) $\ge$ 1.2 (Kewley \&
Dopita, 2002).

	To obtain an estimation of the uncertainties in the abundances,
we have propagated the error in the line fluxes and added
quadratically the accuracy of each method, which is between
$\pm$0.1-0.2 dex. It is worth mentioning that the oxygen abundances
derived in this study correspond to the composition of the ionized
gas phase of the interstellar medium, without taking into account
oxygen depletion in dust grains.

	In this study, we express the chemical abundances as a function
of the galactocentric distance ($r$), related to the Sun as:
[O/H]($r$) = log(O/H)($r$) $-$ log(O/H)$_\odot$, where (O/H)
represents the ratio of the abundance by number of oxygen and of
hydrogen, and (O/H)$_\odot$ is this ratio in the Sun, which has a
value of $-3.34$ dex (Grevesse, Asplund \& Sauval 2007). Moreover,
the radial abundance gradient is expressed by the slope and
y-intercept value of a linear relationship that fits the trend of
empirical abundances as a function of r.

	The Galactocentric distances of the objects have been derived
taking into account the Galactocentric distances derived by Blair et
al. (1982) and Galarza et al. (1999), recomputed for an inclination
angle, $i=77\,^{\circ}$, and adopting the distance to M31 from
Holland (1998): 783 $\pm$ 30 kpc.

	\begin{figure}[!t]
\includegraphics[width=\columnwidth]{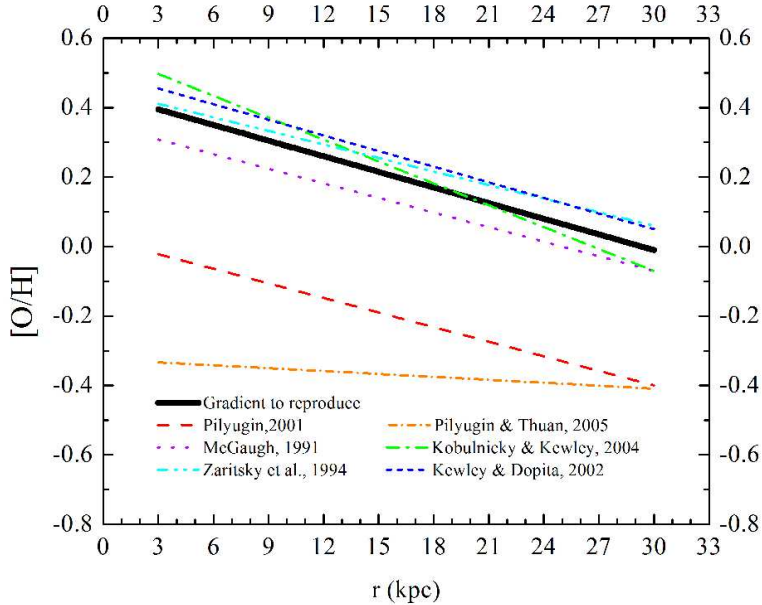} \caption{ Gradients
obtained using different calibration methods. Empirical methods:
Pilyugin (2001) and Pilyugin \& Thuan (2005). Theoretical methods:
McGaugh (1991), Kewley \& Dopita (2002), Kobulnicky \& Kewley (2004),
and Zaritsky et al. (1994). The broad line shows the most probable
gradient, computed applying the method of Akritas \& Bershady (1996)
(see section 2.1.2). This gradient should be reproduced by the
chemical evolution model. Note that the theoretical calibration
methods give very similar gradients.
} 
	\label{fig:gradient} \end{figure}

	In Figure 1, we compile the radial [O/H] gradients obtained from
the different empirical and theoretical calibrations. It is clear
that all the theoretical methods give very similar results in
contrast to the empirical methods, which have much lower y-intercept
values even though they keep similar slopes.

	Additionally, we tested the effect of the presence of intrinsic
scatter in the data in a similar manner to Rosolowsky \& Simon's
(2008) method for the gradient of M33 (the third spiral galaxy of the
Local Group). Following these authors' procedure, we applied the
method of Akritas \& Bershady (1996) to compute the gradients in the
presence of an intrinsic scatter. We constructed histograms for each
calibration taking into account samples of ten HII regions drawn
randomly and we obtained the distributions shown in Figure 2. The
final adopted gradient is the weighted average of the four
theoretical calibrations we have considered. Hence, the adopted slope
is $-0.015 \pm 0.003$ dex kpc$^{-1}$ and the y-intercept value of
[O/H] at the center of M31 is $0.44 \pm 0.04$ dex. Finally, we
adopted two gradients with which to work: at first, the gradient
obtained as the most probable value using the method of Akritas \&
Bershady (1996) ([O/H]($r$) = $-$0.015 $\pm$ 0.003 dex kpc$^{-1}
\times$ r(kpc) + 0.44 $\pm$ 0.04 dex), which we considered to be
representative of the gradients obtained by using photoionization
model calibrations. On the other hand, we tried to explore the
chemical evolution of M31 by adopting the gradient given by
Pilyugin's (2001) empirical calibration, [O/H]($r$) = $-$0.014 $\pm$
0.004 dex kpc$^{-1} \times$ r(kpc) + 0.02 $\pm$ 0.05 dex), a gradient
with similar slope but with a y-intercept value 0.42 dex lower
compared to the most probable gradient obtained from theoretical
calibrations. We also tested the gradient given by Pilyugin \&
Thuan's (2005) empirical calibration, which gives a slope 0.006 dex
kpc$^{-1}$ flatter than all the other calibrations with an
intermediate y-intercept value (see section 4.1 for a more detailed
discussion).

	\begin{figure}[!t]
\includegraphics[width=\columnwidth]{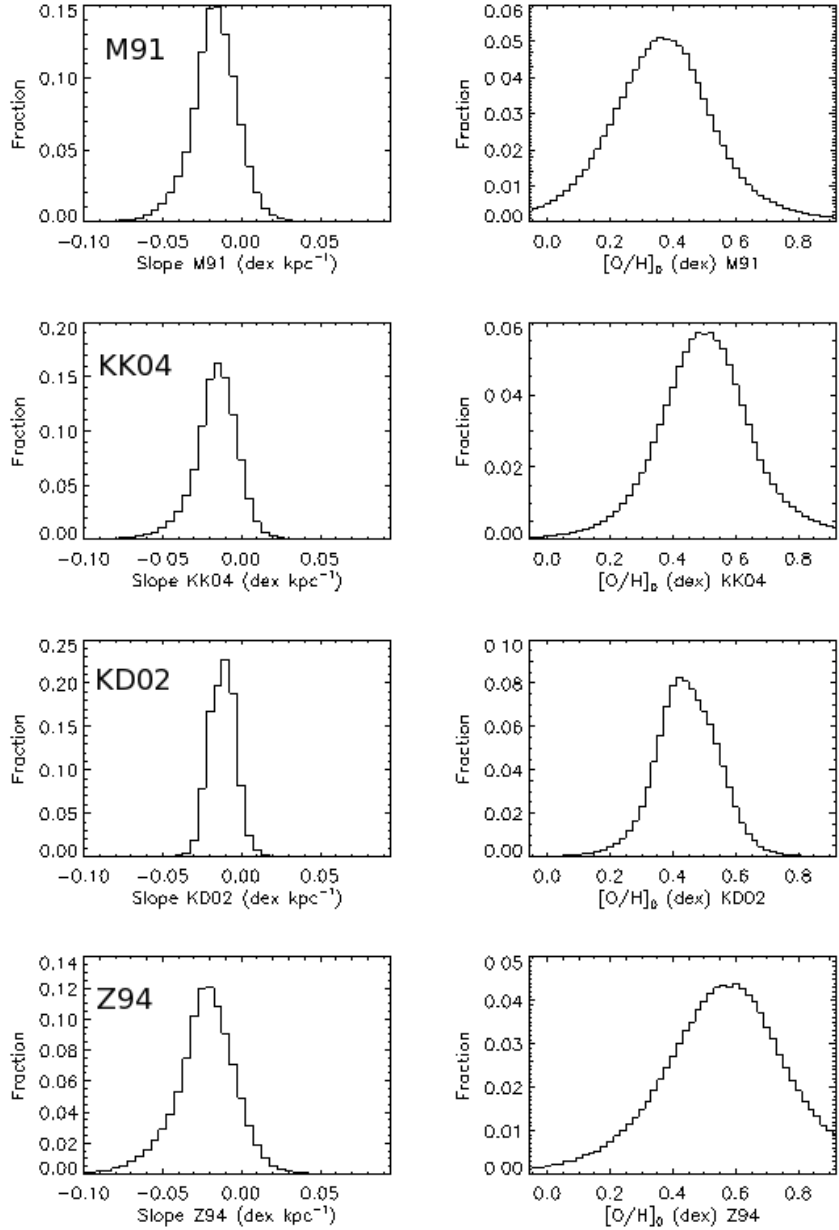} \caption{
Distribution of the slope (left) and y-intercept (right) of the O/H
gradients obtained from the four theoretical calibrations considered
in this work: McGaugh (1991) (M91), Kobulnicky \& Kewley (2004)
(KK04), Kewley \& Dopita (2002) (KD02) and Zaritsky et al. (1994)
(Z94). The distributions are for samples of ten HII regions drawn
randomly from the total sample (see text). See figure 5 of Rosolowsky
\& Simon (2008) for a similar diagram of M33.
} 
	\label{fig:statistics} \end{figure}

%
	\subsubsection{Radial distribution of the gas mass surface
density, $\Sigma_{gas}(r)$} \label{sec:mgas}

	 $\Sigma_{gas}(r)$ represents all gaseous stages which contain
 hydrogen, helium, and the rest of the chemical elements, i.e. 
 $\Sigma_{gas}(r) =  \Sigma_H +  \Sigma_{He} +  \Sigma_Z$.

	E. M. Berkhuijsen (2008, private communication), based on Nieten
et al. (2006), kindly provided us with the updated atomic and
molecular surface density of hydrogen for the northern and southern
halves of the disk of M31. We averaged out these data to obtain
$\Sigma_H$. Taking into account the He and O enrichment by Carigi \&
Peimbert (2008) ($He$ = 0.25 + 3.3 $\times  \ O$, where $He$ and $O$
are abundances by mass), the [O/H] gradient shown in the previous
subsection, and scaling $O$ and $Z$ to solar values, we computed
$\Sigma_{He}$ and $\Sigma_Z$. As additional $\Sigma_H$ data, we used
the compilation of Renda et al. (2005), which was corrected following
the previously described procedure.

	In Figure 3a we show both sets of corrected data which we used as
$\Sigma_{gas}(r)$ constraints on our chemical evolution model.

%
	\subsection{Model's assumptions} \label{sec:ingredients}

	In the present article, we built a dual-infall model in an
inside-out formation framework (i.e., the galaxy is formed more
efficiently in the inner regions than at the periphery), similar to
the model used by Renda et al. (2005), Hughes et al. (2008), and
Carigi \& Peimbert (2011) based on the following assumptions:

	\begin{enumerate}

	\item  The halo and the disk are artificially projected onto
a two dimensional disk with azimuthal symmetry; therefore, all
functions depend only on the galactocentric distance ($r$) and time
($t$).

	\item M31 was formed by a dual-infall, $d\Sigma_T(r,t)/dt$, of
primordial material ($H=0.75$ and $He= 0.25$, which are the
abundances by mass of hydrogen and helium, respectively) given by the
following expression:

	$d\Sigma_T(r,t)/dt = a_h(r) e^{-t/\tau_h} + a_d(r) e^{-(t-1{\rm
Gy})/\tau_d}$.

	The first term represents the halo formation during the first
gigayear (Gy), $a_h(r)$ was obtained considering the halo's
present-day total surface density profile as $ 6 M_\odot {\rm
pc}^2/(1+(r/8 {\rm kpc})^2)$ and $\tau_h = 0.1$ Gy according to Renda
et al. (2005).

	The second term represents the disk formation from 1 Gy until
$t=13$ Gy (current time), where $a_d(r)$ was obtained from
$\Sigma_{T}(r)$ (see section 2.1.1), $\tau_d$ = 0.45 (r/kpc) Gy,
which represents the inside-out scenario and such a value was adopted
to reproduce the $\Sigma_{gas}(r)$ along with [O/H]($r$) (see
sections 2.1.2 and 2.1.3).

	\item The star formation rate (SFR) is the amount of gas which is
transformed into stars and was parameterized as the
Kennicutt--Schmidt law (Kennicutt, 1998), ${\rm SFR} (r,t) = \nu
\Sigma_{gas}^n(r,t)$, where $\nu$ is the efficiency of star formation
and $n$ is a number between 1 and 2. During the halo phase, we chose
$n = 1.0$ and we obtained $\nu= 0.50$ Gy$^{-1}$ in order to reproduce
the maximum average of metallicity shown by the halo stellar
population, log($Z/Z_\odot) = -0.5$ dex (Koch et al. 2008) ($Z_\odot=
0.012$ from Grevesse, Asplund \& Sauval, 2007). For the disk phase,
we chose $n=1.45$, a typical value for the disk of spiral galaxies
(Fuchs, Jahrei$\beta$ \& Flynn 2009), and we obtained $\nu=0.23$
Gy$^{-1}$ (M$_\odot/{\rm pc}^2)^{-0.45}$ in order to reproduce
$\Sigma_{gas}(r)$ along with [O/H]($r$) restrictions at the present
time.

	\item The initial mass function (IMF) is the mass distribution of
the stars formed. We considered the IMF of Kroupa et al.\ (1993),
between a stellar mass interval of 0.1 to 80 M$_\odot$, since the
chemical evolution models that assume this IMF successfully reproduce
the chemical properties of the MW's halo--disk (Carigi et al. 2005) .

	\item We adopted the instantaneous recycle approximation, which
assumes that stars whose initial mass is higher than 1 M$_\odot$ die
instantly after being created and the chemical elements they produce
are ejected into the interstellar medium. This approximation is good
for the elements produced mainly by massive stars, such as oxygen,
since those stars have short lifetimes ($10^{-3}$ to $10^{-2}$ Gy).
The stellar mass fraction returned to the ISM by the stellar
population and chemical yields are taken from Franco \& Carigi
(2008).

	\item The loss of gas and stars from the galaxy to the
intergalactic medium was not considered. All chemical evolution
models of spiral galaxies reproduce the observational constraints
without assuming material loss from the galactic disk (e.g. Carigi \&
Peimbert, 2011). Moreover, no substantial amount of gas surrounding
either M31 or the MW has been observed.

	\item We do not include radial flows of gas or stars.

	\end{enumerate}

%
	\subsection{Results of the Chemical Evolution Model}
\label{sec:results}

	Based on the previously mentioned assumptions and the proper
values of the free parameters chosen in order to reproduce
$\Sigma_T(r,t)$, $\Sigma_{gas}(r,t)$ and the most probable [O/H]
gradient (see section 2.1) at the present time (13 Gy), we obtained the
following results:

%
	\subsubsection{Radial distribution of the gas mass surface
density, $\Sigma_{gas}$} \label{sec:resmgas}

	$\Sigma_{gas}$ was solved numerically for each radius at
different times. With the halo and disk prescriptions shown in
section 2.2, the predicted $\Sigma_{gas}$ at the present time was
found and depicted in Figure 3a along with the data. The maximum of
the theoretical $\Sigma_{gas}$ is shifted towards the inner radius
compared to the observed values, but the agreement with the central
and outer radius is quite good.

	We could improve the $\Sigma_{gas}$ agreement for $r < 8$ kpc,
but the slope of the chemical gradient (the most important constraint
in this study) became steeper than the most probable one and it was
not precise enough to reproduce the chemistry.

	To improve the agreement of  $\Sigma_{gas}$  for $r < 8$ kpc it
may be necessary to assume gas flows through the disk towards the
galactic center, as an effect of the Galactic bar (Portinari \&
Chiosi 2000, Spitoni \& Matteucci 2011), which is beyond the scope of
this paper. Recently, Spitoni et al. (2013) studied the effects of
the radial gas inflows on the O/H gradient of M31. They conclude that
the inside-out galactic formation is the most important physical
process to create chemical gradients, and the radial flows could be a
secondary process. Unfortunately, the effects of the radial gas
inflows on the $\Sigma_{gas}$ are not shown.

	It is difficult to quantify the implications of  the
disagreement between observations and model about $\Sigma_{gas}$ at $r
< 8$kpc, when the other observational constraints are reproduced.
Robles-Valdez, Carigi \& Peimbert (2013) have found a much better
agreement from a more complex chemical evolution model for M31,
assuming the lifetimes of each formed star, a more efficient
inside-out scenario and $r$-dependent $\nu$ (Tabatabaei \&
Berkhuijsen 2010, Ford et al. 2013).

	\begin{figure*}[!t]
\includegraphics[width=0.52\linewidth,height=6cm]{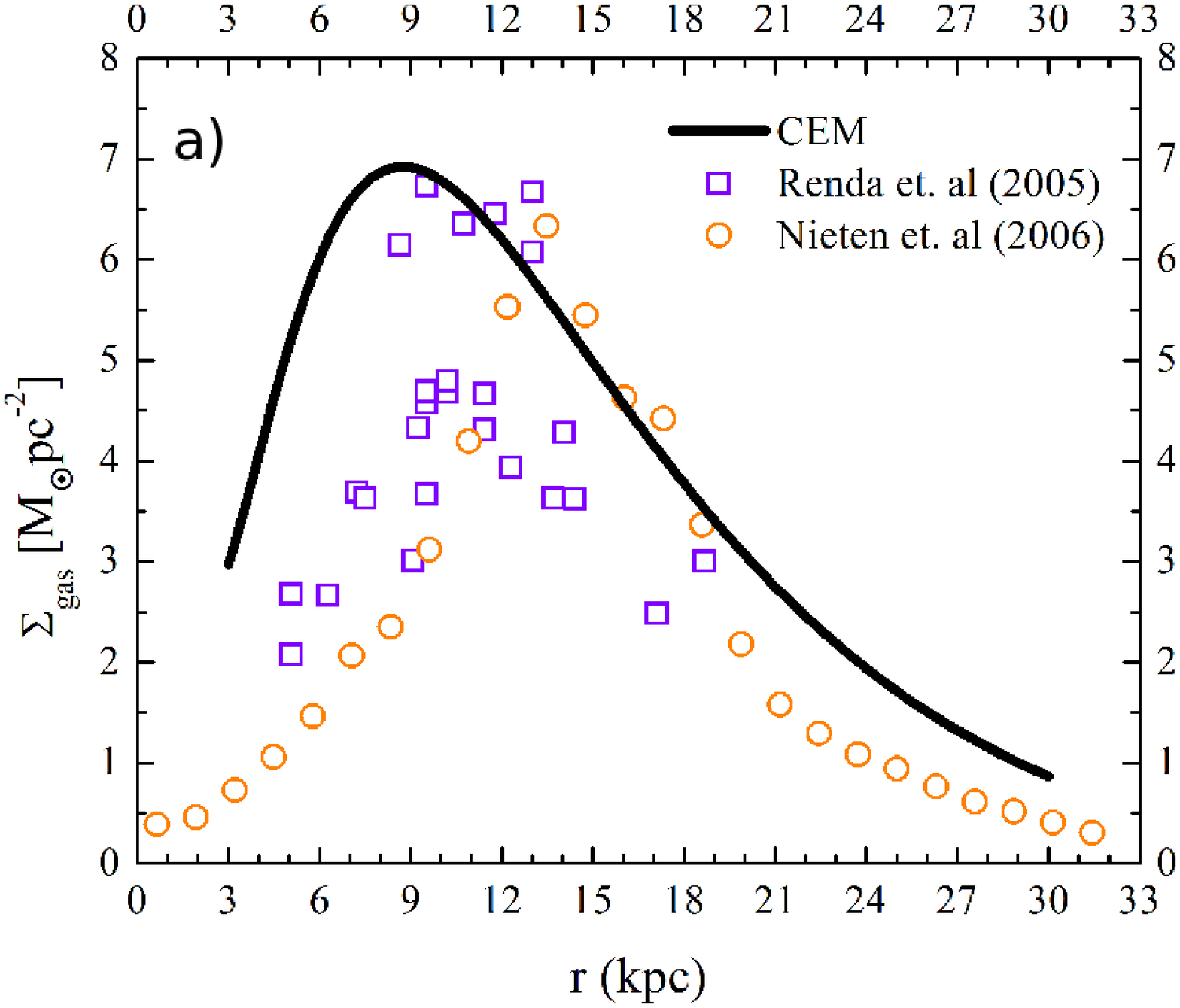}%
\hfill
\includegraphics[width=0.52\linewidth,height=6cm]{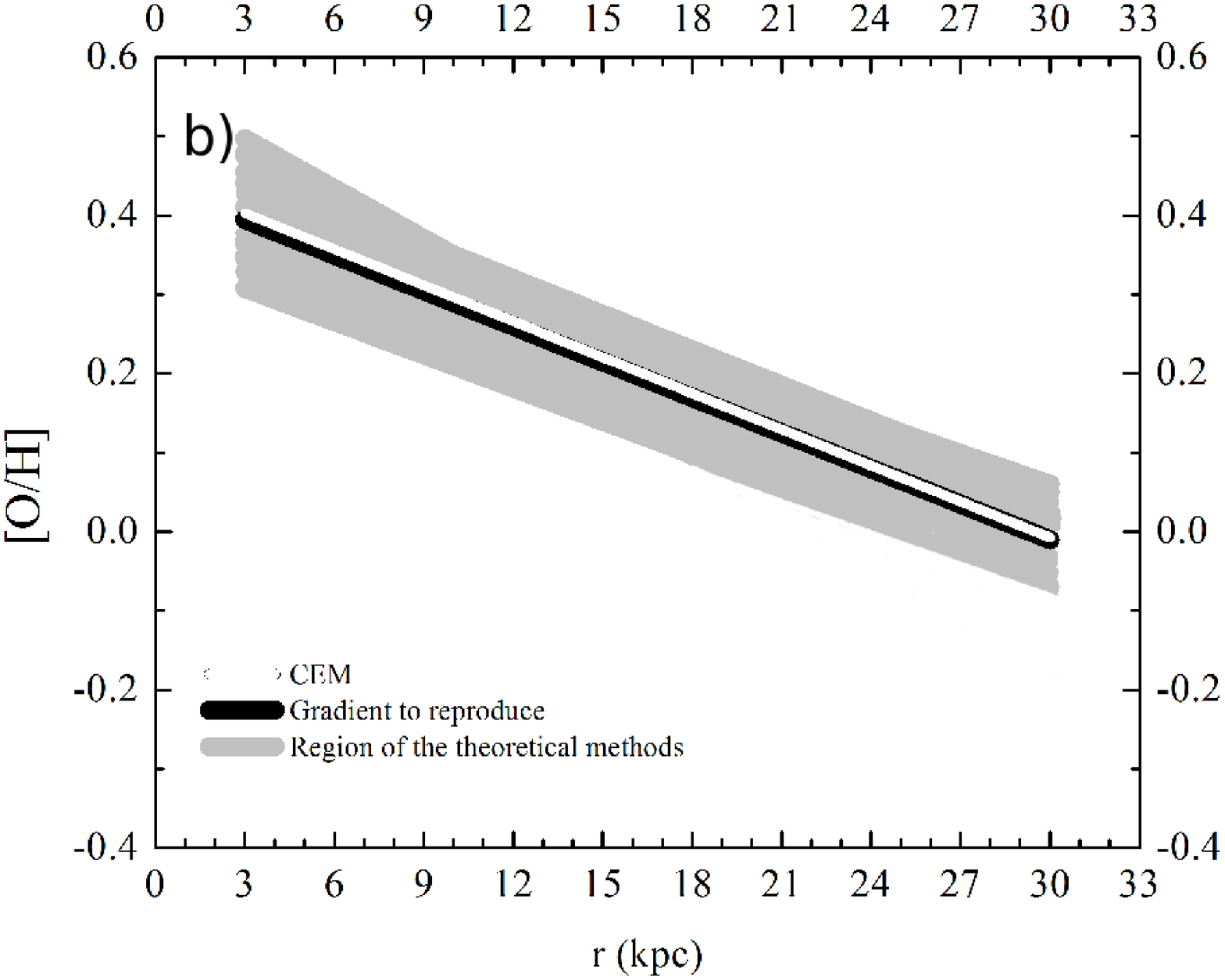}
\caption{ Results of the chemical evolution model (CEM) at the
present-time compared to the observational data. a) Radial
distribution of the gas mass surface density (see section 2.1.3). b)
[O/H] gradient. The white line is the predicted gradient; the black
line represents the most probable gradient and the shaded area
compiles the gradients obtained by the theoretical methods (see
section 2.1.2).
} 
	\label{fig:cem} \end{figure*}

%
	\subsubsection{Chemical abundances} \label{sec:resoh}

	The CEM was built to mainly reproduce the O/H gradient determined
in this study. That gradient allowed us to obtain a reliable chemical
history for M31, where the GHZ will be supported.

	In Figure 3b we present the [O/H] gradient predicted by the model
at the present time (13 Gy)  compared with the observational
constraints for M31. The gradients obtained by the theoretical
methods were enclosed in the shaded area of the figure, and the
adopted gradient, [O/H] = $-$0.015 dex kpc$^{-1} \times$  r (kpc) +
0.44 dex, was also included in the figure as a broad white line. The
gradient obtained from the CEM, [O/H] = $-$0.015 dex kpc$^{-1}
\times$ r(kpc) + 0.45 dex, represented by a solid black line in the
figure, is in perfect agreement with the observed value, ensuring the
predicted metallicity history.

	As previously described, oxygen is the most abundant elemental
component of $Z$; a chemical evolution model built to reproduce the
[O/H] gradient therefore allows an adequate approximation of the
evolution of the heavy elements, represented by $Z$.

	Since the GHZ depends mainly on metallicity, we are interested in
the evolution of the metal radial distribution; hence, in Figure 4a
we show the behavior of log($Z/Z_\odot$) with respect to
galactocentric distances at different times. It is not surprising
that the current gradient of log($Z/Z_\odot$), log($Z/Z_\odot$) $=
-0.015$ dex kpc$^{-1} \times$ r (kpc) + 0.41 dex, is similar to the
[O/H] gradient obtained by the CEM, [O/H] = $-$0.015 dex kpc$^{-1}
\times$ r (kpc) + 0.45 dex, confirming that oxygen is the most
abundant heavy element and behaves as $Z$.

	From Figure 4a, we notice that the gradient flattens from 3 to 13
Gy, due to the inside-out scenario: at the beginning of the evolution
the infall was relevant in the central regions compared to the outer
parts, producing higher $\Sigma_{gas}$ and SFR, rapidly increasing
the oxygen abundance in the inner regions; then the infall, and
therefore the SFR, dropped, causing a lower increase in the oxygen
abundance. In contrast, at the final stages of the evolution, the
infall became important in the outer regions compared to the inner
regions, thus $\Sigma_{gas}$ increased and SFR became more efficient,
producing a higher increase in the O abundance in outer regions and,
consequently, the gradient became flatter.

	The positive slope of the $Z$ gradient at 1 Gy is quite
remarkable. This opposite slope behavior is due to the enormous
amount of primordial material, from the intergalactic medium, that
fell into the inner parts at the beginning of the disk formation,
causing gas dilution and a consequent decrease in the O/H
values.

	\begin{figure*}[!t]
\includegraphics[width=0.52\linewidth,height=6cm]{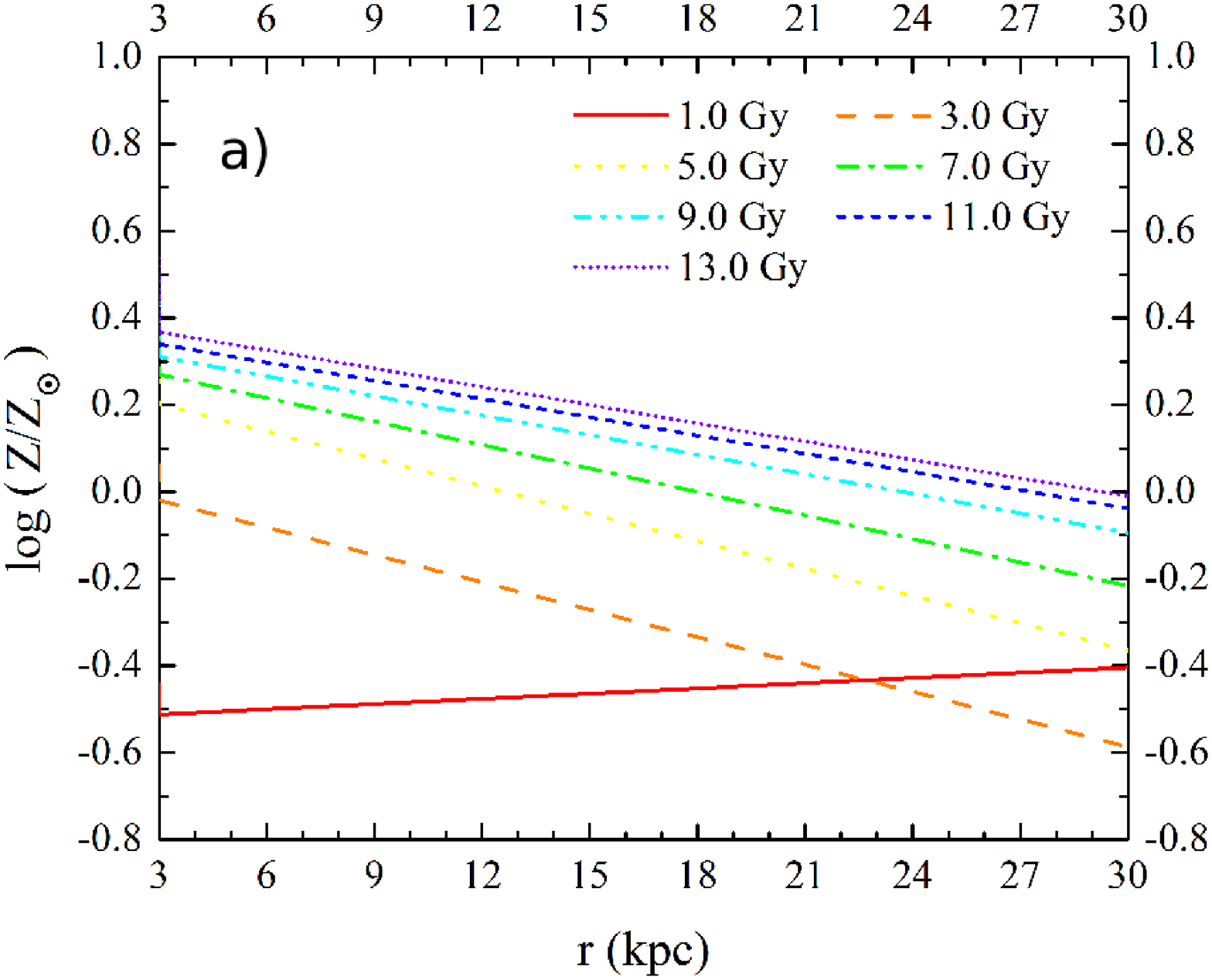}%
\hfill
\includegraphics[width=0.52\linewidth,height=6cm]{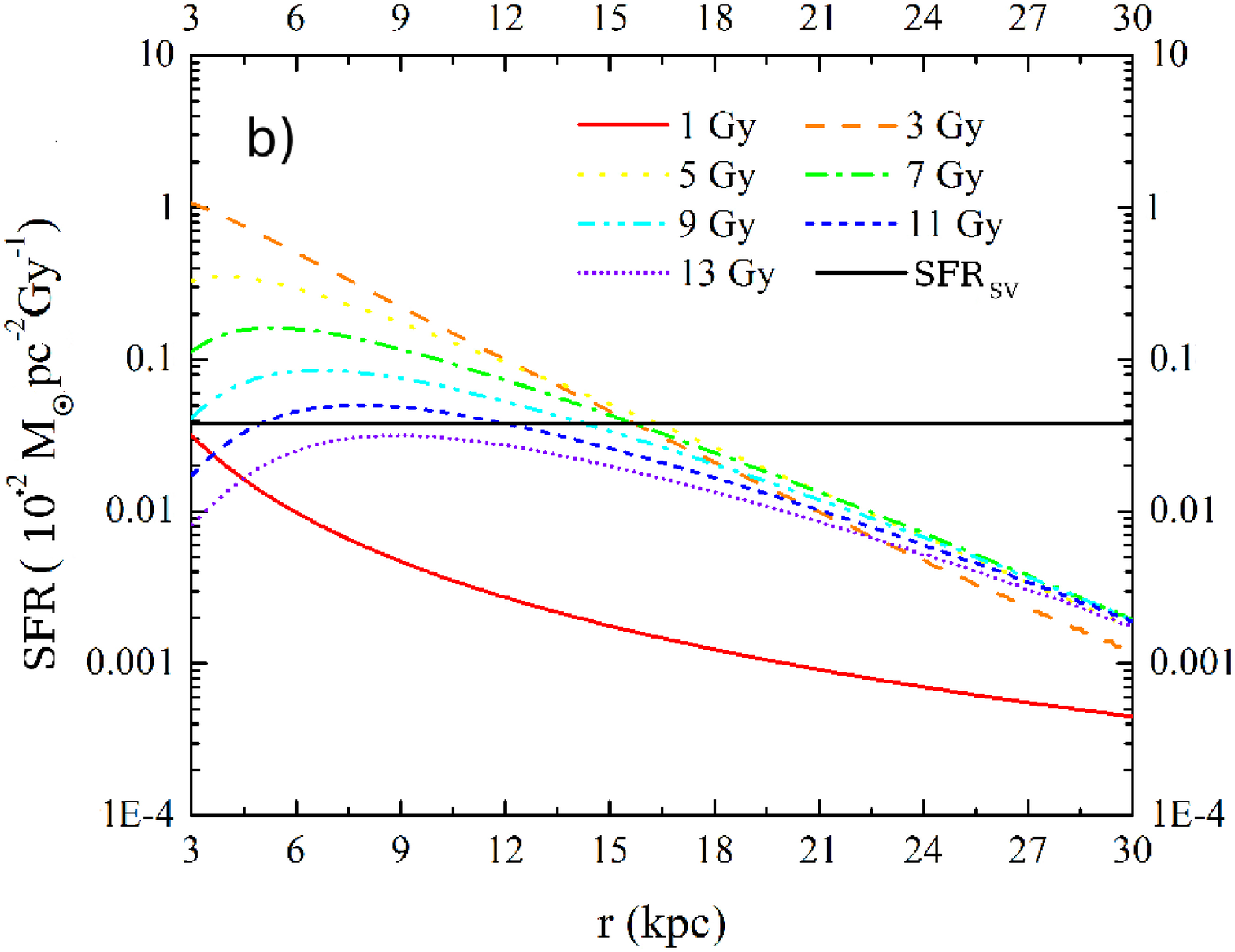}
\caption{ Evolution of radial distributions at 1, 3, 5, 7, 9, 11, and
13 Gy. a) Metallicity relative to the solar value, log($Z/Z_\odot$)
($Z_\odot= 0.012$, Grevesse et al. 2007). b) Star formation rate,
SFR. The horizontal line represents the average value of the SFR that
the solar neighborhood has undergone during the Sun's lifetime.
(Carigi \& Peimbert 2008).
} 
	\label{fig:evol} \end{figure*}

%
	\subsubsection{Supernova rate} \label{sec:ressn}

	As we have mentioned, the occurrence of SN is
important in the calculation of the GHZ because a high SN rate may
 deplete the ozone layer in a planetary atmosphere and,
thereby, life. Since our CEMs is built using the instantaneous
recycle approximation, we cannot estimate the SNIa and consequently,
they are not included in the SN rate. That is a model limitation,
particularly at high galactic times, because the time-delay of a SNIa
to pollute the gas is between 0.1 Gy and the Hubble time (Maoz et al.
2012). That limitation would be counterbalanced by the low average of
SNIa in Sbc/d galaxies, like M31 (SNIa/SNII$ =0.17$, Mannucci et al.
2005).
	From the star formation rate inferred by our chemical evolution
model, we obtained the SN rate. It was computed according to 
$RSN(r,t) = N_{MS} \times {\rm SFR}(r,t)$, where $N_{MS} = 0.05
M_\odot^{-1}$ is the number of type II supernova progenitors per
$M_\odot$. We assumed that those progenitors are stars more massive
than 8 $M_\odot$ and therefore $N_{MS}$ was calculated by integrating
the initial mass function between 8 and 80 $M_\odot$.

	In Figure 4b, we show the star formation rate as a function of
galactocentric distance for the same times shown in Figure 4a. In
order to compare with the SFR that occurred in the solar neighborhood
during the Sun's lifetime, we added the average value of the SFR in
the solar neighborhood ($r=8$ kpc) over the last 4.5 Gy. That average
was obtained from the mean SFR at the solar radius during the last
4.5 Gy, $<$SFR (8 kpc)$> = <$SFR$_{SV}> = $  3.8 M$_\odot$ Gy$^{-1}$
pc$^{-2}$ (see Fig. 2 by Carigi \& Peimbert 2008). Consequently, the
4.5 Gy average SN rate of the solar neighborhood during the
Earth's age is $<RSN_{SV}> = 0.2 \ Gy^{-1}$ pc$^{-2}$.

	\section{Galactic Habitable Zone} \label{sec:ghz}

	The Galactic Habitable Zone (GHZ), is defined as the region with
sufficient abundance of chemical elements to form planetary systems
in which Earth-like planets could be found and might be capable of
sustaining life. Therefore, a minimum metallicity is needed for
planetary formation, which would include the formation of a planet
with Earth-like characteristics (Gonzalez et al. 2001, Lineweaver
2001), and a CEM provided the evolution of the heavy element
distribution (see previous section).

	\subsection{Characteristics and conditions for the GHZ}
\label{sec:propghz}

	In this context, it is important to bear in mind the properties
and characteristics that an Earth-like planet and its environment
need in order to be considered habitable or leading to habitability.
In astrobiology, the chemical elements are classified according to
the roles they play in the formation of an Earth-like planet and in
the origin of life: the major biogenic elements are those that form
amino acids and proteins (e.g. H, N, C, O, P, S), the geophysical
elements are those that form the crust, mantle, and core (e.g. Si,
Mg, Fe) (Hazen et al. 2002).

%
	\subsubsection{Astrophysical conditions} \label{sec:astro}

	The conditions of the environment around an Earth-like planet are
the following:

	\begin{enumerate}
	
	 \item Planet formation: The metallicity in the
medium where the planet may form (protoplanetary disk) must be such
that it allows matter condensation and hence protoplanet formation
(Lineweaver 2001). Therefore, Earth-like and Jupiter-like planets may
be created. Whether the creation of Jupiter-like planets is
beneficial or not is a debatable subject (Horner \& Jones 2008,
2009). They may act as a shield against meteoritic impact (Fogg \&
Nelson 2007, Ward \& Brownlee 2000), or might migrate towards the
central star of the planetary system, thereby affecting the internal
planets, or even destroying them (Lineweaver 2001, Lineweaver et al.
2004). On the other hand, the simulation by Raymond et al. (2006)
predicts that Earth-like planets may form from surviving material
outside the giant planet's orbit, often in the habitable zone and
with low orbital eccentricities. Most of the planet-harboring
stars, detected by several research teams, show a wide range of
stellar metallicity, $-0.76 \le {\rm log}(Z/Z_\odot) \le +0.56$, and
the distribution peaks at log($Z/Z_\odot) \sim +0.20$ (See Figure 5,
lower pannel). Data compilation taken from The Extrasolar Planets
Encyclopaedia on March 2013, http://exoplanet.eu/catalog.php).
Moreover, based on the same data source, the observed relation
between stellar $Z$ abundance and the planet mass presents a wide
dispersion. Since that sample includes all types of planets---both
Earth-like and Jupiter-like---the stellar $Z -$planet mass relation
does not show a clear trend. Recently, Jenkins et al. (2013) have
grown the population of low-mass planets around metal-rich stars,
increasing the dispersion in the stellar $Z -$planet mass relation.
On the other hand, Adibekyan et al. (2012) confirm an
overabundance in giant-planet host stars with high $Z$. Therefore, we
think that the fraction of detected exoplanets as a function of the
metallicity of stars they orbit, represents the necessary $Z$ to form
planets that survive to planetary migrations. It should be mentioned
that the stellar $Z$ is associated with the planet $Z$ since the
planets were created from the same gaseous nebula of the star.
 Future studies will take into account planet occurrence from {\it Kepler} mission
(e.g.  see Howard et al. 2012).

	\item Survival from SN: When a supernova explodes, emits
strong radiation that may ionize the planet's atmosphere,
causing the stratospheric-ozone depletion. Then ultraviolet flux
from the planet's host star reaches the surface and oceans,
originating damage to genetic material DNA, which could induce
mutation or cell death, and consequently the planet sterilization
(Gehrels et al., 2003). Since the Earth is the only known planet with
life, we assumed it could represent the survival pattern to SN
explosions. We have analyzed three possibilities for the planet
survival, such that the life on the planet will be annihilated
forever if: i) the instantaneous $RSN(r,t)$ is higher than
$<RSN_{SV}>$, ii)  the average $RSN$, during the entire existence of
the planet, is higher than $<RSN_{SV}>$, and iii)  the average $RSN$,
during the first 4.5 Gy of the planet's age, is higher than $ 2
\times <RSN_{SV}>$, being $<RSN_{SV}> = 0.2  \ {\rm Gy}^{-1} {\rm
pc}^{-2}$ the time-average of SN rate in the solar neighborhood,
during the last 4.5 Gy. In all our RSN computations we have not
included SNIa and we could be undervaluing the amount of SN
that may sterilize planets, mainly at recent times. We think that
this underestimation is compensated for by an overestimation of
SNII, due to the instantaneous recycle approximation, and by the low
fraction of SNIa ($\sim$ 2 SNIa per 10 SNII) for galaxies like M31
(see section 2.3.3).

	\end{enumerate}

	\subsubsection{Geophysical characteristics} \label{sec:geo}

	An Earth-like planet is one that has enough biogenic and
geophysical elements to sustain and allow the development of life as
it is known on Earth (Sleep, Bird \& Pope 2012, Bada 2004,
Gomez-Caballero \& Pantoja-Alor 2003, Hazen et al. 2002, Orgel 1998).
In general terms, for a planet to be identified as an Earth-like
planet,  it has to satisfy the following conditions:

	\begin{enumerate}

	\item Tectonic plates: a crust formed mainly  of Si constitutes
the tectonic plates. These are necessary since life must have a site
to live on with enough resources for survival. The recycling of the
tectonic plates keeps the density and the temperature of the
planetary atmosphere and the amount of liquid water and carbon needed
for life. (Hazen et al. 2002, Lineweaver 2001, Segura \& Kaltenegger
2006 ).

	\item Water: on the surface of the tectonic plates are found the
oceans, formed by H$_2$O. Since life on Earth is the pattern for
life, water is a crucial resource for the emergence and survival of
life (McClendon 1999).

	\item Atmosphere: the atmosphere of a planet should be dense
enough to protect the planet from UV radiation and meteoric impacts,
and thin enough to allow the evolution of life on the planet's
surface. The atmospheric composition of the primitive Earth allowed
the origin of life and had abundance of CO, CO$_2$, H$_2$O, N$_2$O,
and NO$_2$. Such compounds were crucial to the origin of the present
atmosphere on Earth. (Bada 2004, Chyba \& Sagan 1991, Maurette et al.
1995, Navarro-Gonz\'alez et al. 2001, Sekine et al. 2003).
\end{enumerate}

	Since all the geophysical elements are heavier than He, the
abundance of these kinds of elements necessary to create an
Earth-like planet is both contained and well represented by the
distribution of $Z$ given by the chemical evolution model.

	\subsubsection{Biogenic characteristics} \label{sec:bio}

	As previously mentioned, the GHZ is based on the pattern of life
on Earth; therefore, in order to determine the time of  the origin
and development of life on Earth several studies were taken into
account.

	There are several theories of the means and places where life
originated on Earth. Some of those theories are based on hydrothermal
vents on the seabed, where an interaction between the terrestrial
mantle and the ocean exists (Bada 2004, Chang 1982, Gomez-Caballero
\& Pantoja-Alor 2003, Hazen et al. 2002, McClendon 1999). Other
theories have assumed the migration of life from other regions of
space to Earth (Maurette et al. 1995, Orgel 1998). There are others
explaining the catalytic effect of lightning or metallic meteorites
(Chyba \& Sagan 1991,  Navarro-Gonz\'alez et al. 2001, Sekine et al.
2003) in the primitive atmosphere and/or in the seas. Regardless of
how  life emerged, all theories have agreed on the same biogenic
requirements, i.e. C, N, O, P, S. Those biogenic elements are heavier
than He and the abundance of these elements is consequently contained
and well represented by the distribution of $Z$ given by the CEM.

	The earliest evidence we have for life on Earth is about 3.5 Gy
ago (the most ancient fossils being cyanobacteria, Kulasooriya
2011, and references therein) or 3.8 Gy (contested carbon isotope
evidence, Mojzsis et al. 1996). There is general agreement that
life got started 3.7 Gy ago (Ricardo \& Szostak 2009). The Earth
is $4.6 \pm 0.1$ Gy old (Bonanno et al. 2002); therefore, life took
around 0.9 Gy to emerge. In consequence, in this study we consider
1.0 Gy  as the minimum age of a planet 
capable of hosting basic life on its surface.

	On the other hand, if evolved life is associated with humans,
another parameter of life should be taken into account since humans
appeared on Earth 2 million years ($2 \times 10^{-3}$ Gy) ago (Bada
2004). However, the fact that the times used in our CE study are of
the order of gigayears, this $2 \times 10^{-3}$ Gy is negligible. 
Thus, in this study we consider 4.5 Gy  as the minimum age of a planet capable of sustaining complex life.

	\subsection{Restrictions for the GHZ} \label{sec:resghz}

	Based on the previously collected information, in order to obtain
the evolution of the Galactic Habitable Zone in M31, we chose the
following astronomical and biogenic restrictions:

	\begin{enumerate}

	\item  Stars might harbor Earth-like planets.

	\item Earth-like planets form from gas with specific
$Z$-dependent probabilities.

	\item Earth-like planets require 1.0 Gy to create basic life
(BL).

	\item Earth-like planets need 4.5 Gy to evolve complex life (CL).

	 \item Life on formed planets is annihilated forever by the
SN explosions under the following conditions: \begin{enumerate} \item
the SN rate at any time and at any radius has been higher than the
average SN rate in the solar neighborhood ($<RSN_{SV}>$) during the
last 4.5 Gy of the Milky Way's life, \item the average $RSN$, during
the whole existence of the planet, is higher than $<RSN_{SV}>$, and
\item the average $RSN$, during the first 4.5 Gy of  the planet life,
is higher than $2 \times <RSN_{SV}>$ \end{enumerate}
	\end{enumerate}

	\begin{figure}[!t]
\includegraphics[width=\columnwidth]{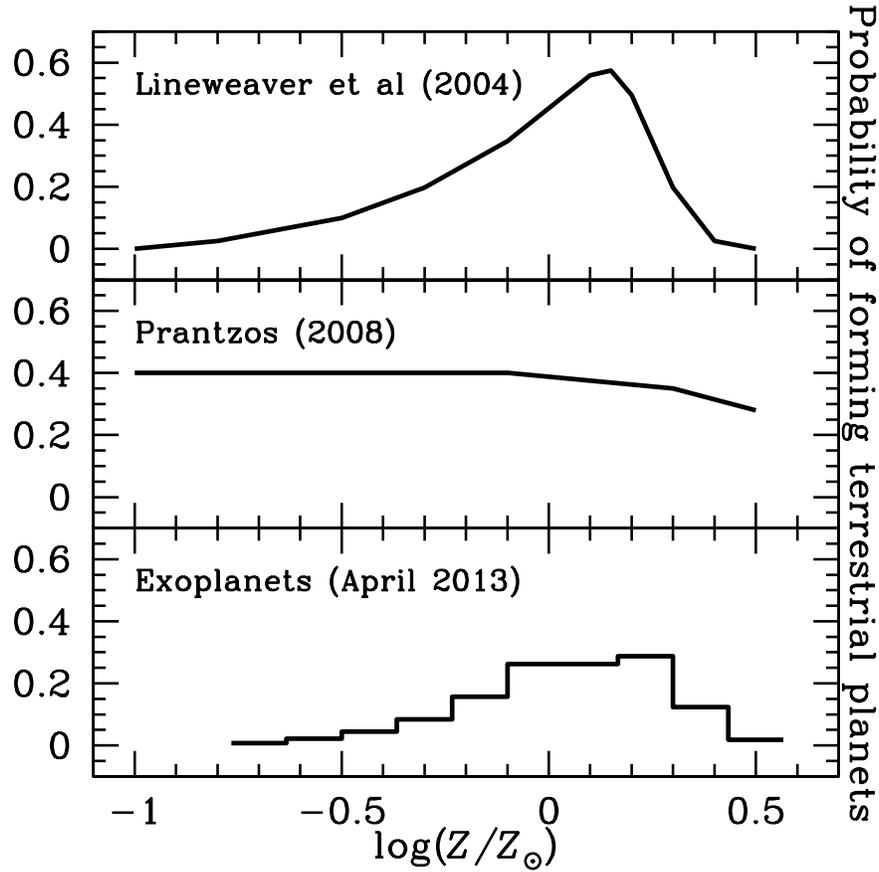} \caption{ 
Probability of terrestrial planet formation by Lineweaver et al.
(2004, upper panel), Prantzos (2008, middle panel) and extrasolar
planets (The Extrasolar Planets Encyclopaedia on March 2013,
http://exoplanet.eu/catalog.php, lower panel).
}
	\label{fig:statistics} \end{figure}

	 Based on the previous assumptions, the probability, $P(r,
t)$, to form Earth-like planets, orbiting stars, which survive  SN
explosions and where basic and complex life may surge is calculated
as:

	$P_{GHZ}(r, t) = P_{STAR}(r,t) \times P_Z(r,t)  \times P_{BL}(t)
\times P_{CL}(t) \times P_{SN}(r,t)$,

	where: 
	\begin{enumerate}

	\item  $P_{STAR}(r,t) = SFR(r,t) dt$ is the probability of
forming new stars, per surface unit ($pc^2$), during a delta time.

	\item  $P_Z(r,t)$ is the $Z$-dependent probability of forming
 terrestrial planets (see Figure 5): \begin{enumerate} \item as
Lineweaver et al (2004) assume, \item as Prantzos (2008) consider,
and \item identical to the $Z$ distribution shown by extrasolar
planets (The Extrasolar Planets Encyclopaedia on March 2013,
http://exoplanet.eu/catalog.php). \end{enumerate}

	\item $P_{BL}(t)$ and $P_{CL}(t)$ are the probabilities of the
emergence of basic life and the evolution of complex life. We adopt
functions with no dispersion (equal to 0.0 or 1.0) because the
uncertainties in the evidence for life are similar to the temporal
resolution of the chemical evolution model.

	\item $ P_{SN}(r,t) $ is the probability of survival of supernova
explosions. We adopt $ P_{SN}(r,t)=1.0 $ or $ P_{SN}(r,t)=0.0 $ if SN
rate is lower or higher than the average $RSN$, respectively. This is
a good approximation because of the quick decay of $P_{SN}$ 
(Lineweaver et al. 2004) and  the two extreme conditions supported on
the pattern of life and considered in this study.

	\end{enumerate}

	\subsection{Results of the Galactic Habitable Zone}
\label{sec:resughz}

	Our chemical evolution model---explained in section 2---provided
the evolution of: metals, star formation rate, and SN rate at
different galactocentric distances, which are the data required to
compute the Galactic Habitable Zone. We then applied the restrictions
listed in the previous sections on the galactic disk of M31 and
obtained GHZs. Since most previous studies of the GHZ have focused on
the Milky Way (Lineweaver et al. 2004, Prantzos 2008, Gowanlock et
al. 2011), we also obtained the GHZ of the Andromeda galaxy using
their common restrictions in order  to compare the GHZ in both
neighboring spiral galaxies using identical galactic habitable
conditions.

	\subsubsection{ SN effects on the GHZ} \label{sec:work}

	Based on the results of the chemical evolution model, we found
the galactic times and regions in the galactic disk where the
metallicity satisfies the exoplanet distribution, then we account for
the number of formed stars and we apply the condition of SN
survival.

	\begin{figure}[!t]
\includegraphics[width=\columnwidth]{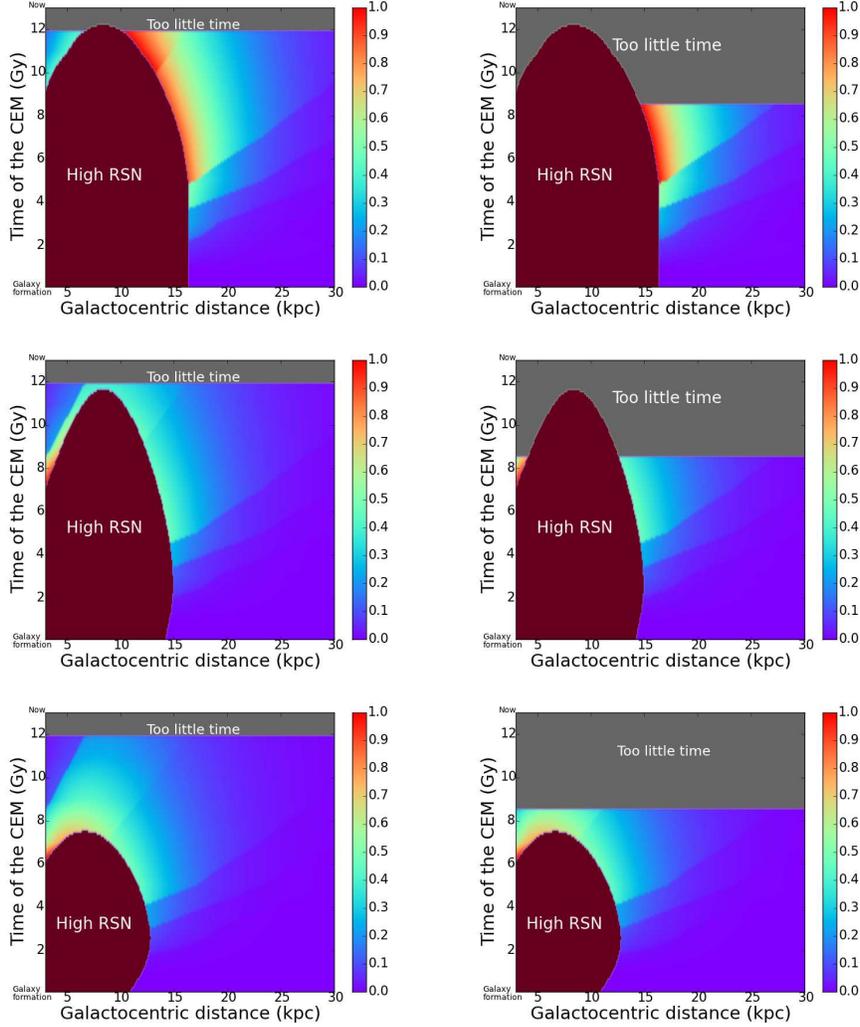} \caption{ 
Evolution of the Galactic Habitable Zone of M31 assuming the $Z$
distribution shown by exoplanets, three different supernova survival
conditions, and two type of life on the exoplanets. Planet are
sterilized if: i)  the instantaneous SN rate is higher than the
time-average SN rate of the solar neighborhood during the Sun's
existence, $<RSN_{SV}>$(upper panels), ii)  the average SN rate
during the planet's entire existence is higher than $<RSN_{SV}>$
(middle panels), and iii) the average $RSN$ during the first 4.5 Gy
of  the planet life is higher than $2 \times <RSN_{SV}>$ (lower
panels). Planets capable ofsustaining basic life (left panels) or evolved life
(right panels). Burgundy area: zone where life may be extinguished by
SN explosions. Grey area: zone where planets may not harbor life. Red
and violet areas: zones with maximum and minimum probabilities,
respectively, to form Earth-like planets orbiting stars. Other
colors:  intermediate probabilities of the GHZ.
} 
	\label{fig:snghz}
	\end{figure}

	 In Figure 6 we show the evolution of the GHZ, per surface
unit, assuming these constraints for three SN survival conditions
; i.~e. the planet sterilization occurs if: i)  the instantaneous SN
rate is higher than $<RSN_{SV}>$ (upper panels), ii)  the average SN
rate during the planet's entire existence ($<RSN(r,t)>_{planet \
age}$) is higher than $<RSN_{SV}>$ (middle panels), or iii) the
average $RSN$ during the first 4.5 Gy of  the planet life
($<RSN(r,t)>_{4.5 Gy}$)  is higher than $2 \times <RSN_{SV}>$ (lower
panels). Moreover, we imposed the restriction of basic life (1 Gy,
left panels) and evolved life (4.5 Gy, right panels).

	We mark where and when planets were sterilized (burgundy) and
where and when planets may not be capable of hosting basic and complex life (grey). We
indicate with differently colors the probability values to form
Earth-like planets that orbit stars. The red and violet areas
represent the zones with maximum and minimum probabilities.

	The GHZs shown in Figure 6 are smaller at high planet ages
(equivalent to low evolutionary time) and excludes short
galactocentric radii, due to the inside-out formation scenario, upon
which the chemical evolution model was built.  That model predicts a
high star formation rate at inner radii during the first moments of
evolution and, consequently, a high SN rate that would have
sterilized the planets formed there.

	If life has not recovered after $RSN(r,t) = <RSN_{SV}>$, the
most likely GHZ is located in the middle galactic disk ($10 - 18$
kpc). But if life has not recovered when $<RSN(r,t)>_{planet \ age} =
<RSN_{SV}>$ the most likely GHZ is located in the very inner galactic
disk ($3 - 5$ kpc).

	Since life on Earth has proven to be highly resistant, we think
that the GHZ assuming $<RSN(r,t)>_{4.5 Gy} = 2 \times <RSN_{SV}>$ is
more reliable. Therefore, based on the lower panels of Figure 6, the
most likely GHZ (per surface unit) is located in the inner disk ($3 -
7$ kpc) with ages between 6 and 7 Gyr. The GHZ with median
probabilities evolves from the middle ($\sim 13$ kpc) to the inner
disk, from 4.5 Gy to 8.0 Gy age. If we consider only basic life, the
younger GHZ evolves away from the inner parts. In either case, the
halo component (for evolution times lower than 1 Gy at any $r$) is
discarded from the GHZ because of its low $SFR$ and $Z$.

	\begin{figure}[!t]
\includegraphics[width=\columnwidth]{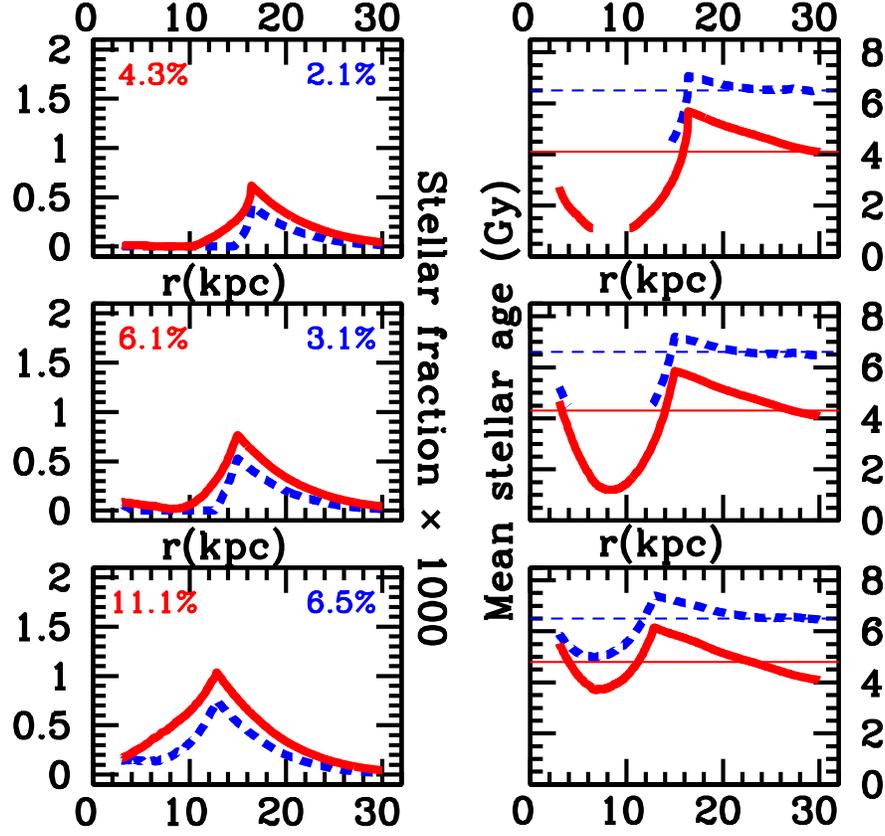} \caption{ Left
panels: $0 - 13$ Gy integrated fraction of stars harboring
terrestrial planets formed according to $Z$ distribution shown by
exoplanets and that survive SN explosions: $RSN(r,t) \leq  \
<RSN_{SV}>$ (upper panels), $<RSN(r,t)>_{planet \ age} \leq  \
<RSN_{SV}>$ (middle panels), and $<RSN(r,t)>_{4.5 Gy} \leq 2 \times
<RSN_{SV}>$ (lower panels). Red-continuous lines: basic life.
Blue-dashed lines: complex life. Left and right corners: total
percentage of planetary systems capable of sustaining basic life or complex life,
respectively, on their Earth-like planets. Right panels: mean age of
the stellar fractions shown in the left panels. Thin lines: the
average age of the total stars with solid planets that survive SN.
} 
	\label{fig:snfraction} \end{figure}

	 In the left panels of Figure 7, for the same GHZs of Figure
6, we plot the fraction of stars with terrestrial planets that
survive SN explosions, taking into account basic and complex life may
surge (red-continuous and blue-dashed lines, respectively). That
fraction represents the $0 - 13$ Gy integrated number of those stars
formed in a d$r-$width ring centered in each $r$, normalized by the
total number of stars formed in the $3-30$ kpc disk. In the left and
right corners, we show the total percentage of stars harboring
Earth-like planets that survive SN and capable of hosting basic life or evolved
life, respectively. Obviously, the percentage with
basic life is higher than the percentage with complex life, and both
percentages increase with the life's resistance to the SN harmful
effects. It is important to note that the peak in the stellar
fractions are not located at the same radius as the GHZs due to
the time-integration of the $P_{GHZ}(r, t)$ and  the total area of
each ring.

	In the right panels of Figure 7 we plot the mean age of the
stellar fractions shown in the left panels. In the upper and middle
windows there are not data for some radii less than 15 kpc, 
due to all terrestrial planets formed at those $r$ were sterilized by SN.

	For $r < 13-15$ kpc, the stellar fraction and the mean stellar
age decrease with the increase of SN effects. For $r > 15$ kpc, the
$r$ behavior of the stellar fraction and the mean stellar age are
identical between GHZs because the SN effects are null and the
$P_{STAR}(r,t)$ and $P_Z(r,t)$ are identical. The thin lines
represent the average age of the total stars harboring Earth-like
planets that survive SN. For complex life and any SN survival
conditions,  the average stellar age is $\sim 6.5$ Gyr, but for basic
life the average age is $4 - 5$ Gy and increasing with life's
resistance to the SN effects.

	\subsubsection{ $Z$ effects on the GHZ} \label{sec:mw}

	Four of the five previous studies of the GHZ (Gonzalez et al.
2001, Lineweaver et al. 2004, Prantzos 2008, Gowanlock et al. 2011,
Suthar \& McKay 2012) have focused on the Milky Way.

	Lineweaver et al. (2004), Prantzos (2008) and Gowanlock et al.
(2011) have computed the GHZ of the MW assuming specific $Z$ ranges
to form Earth-like planets and for SN survival. Since Gowanlock et
al. adopted a much wider $Z$ range than that of Lineweaver et al.,
with probabilities lower and almost $Z$ independent, similar to
that by Prantzos, we computed the GHZ of M31 based on the assumptions
of Lineweaver et al. and Prantzos, separately (Figure 5, upper and
middle panels). Moreover we considered that life is annihilated
forever when $<RSN(r,t)>$, during the first 4.5 Gy of  the planet, is
higher than $2 \times <RSN_{SV}>$, SN condition similar to that used
by Lineweaver et al. and Prantzos.

	\begin{figure*}[!t]
\includegraphics[width=\columnwidth]{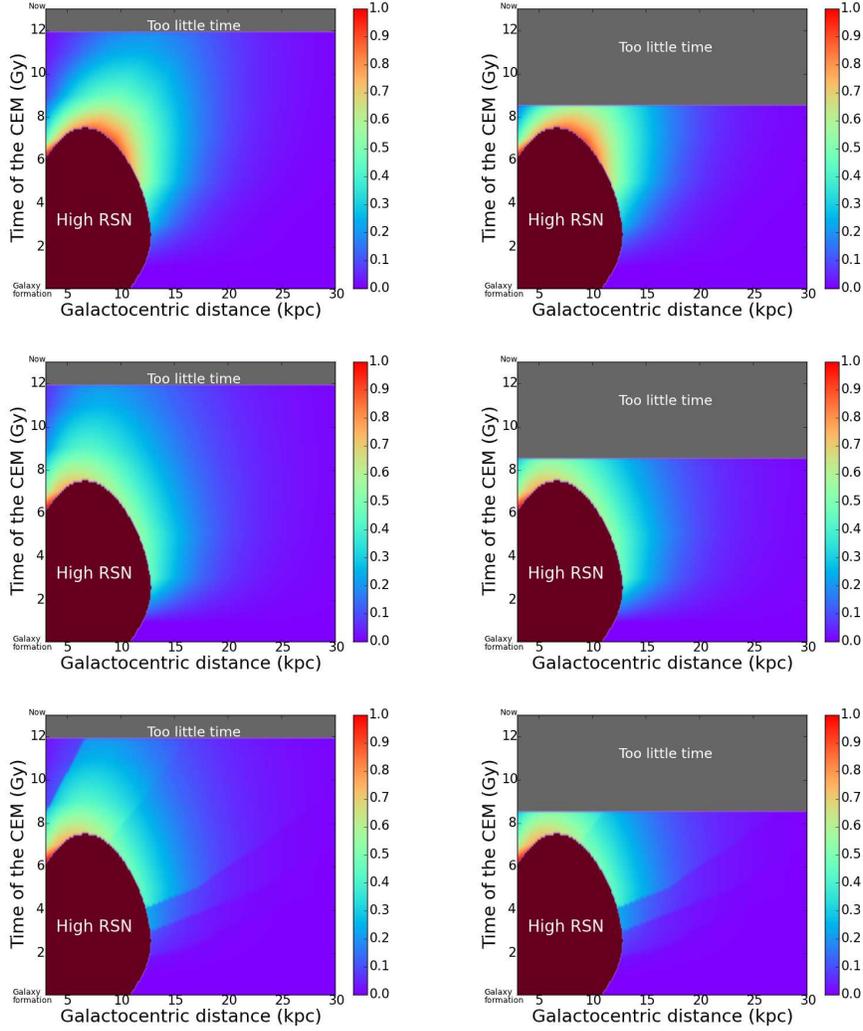} \caption{
Evolution of the Galactic Habitable Zone of M31 assuming that planet
sterilization occurs when the average $RSN$ during the first 4.5 Gy
of the planet is higher than $2 \times <RSN_{SV}>$ and three
different $Z$ probability of forming terrestrial planets (see Figure
5) by: i)  Lineweaver et al. (2004, upper panels), ii)  Prantzos
(2008, middle panels), and iii) extrasolar planets (2013, lower
panels). Planets capable of hosting basic life (left panels) or evolved life
(right panels). Colors as Figure 6.
} 
	\label{fig:snghz} \end{figure*}

	\begin{figure}[!t]
\includegraphics[width=\columnwidth]{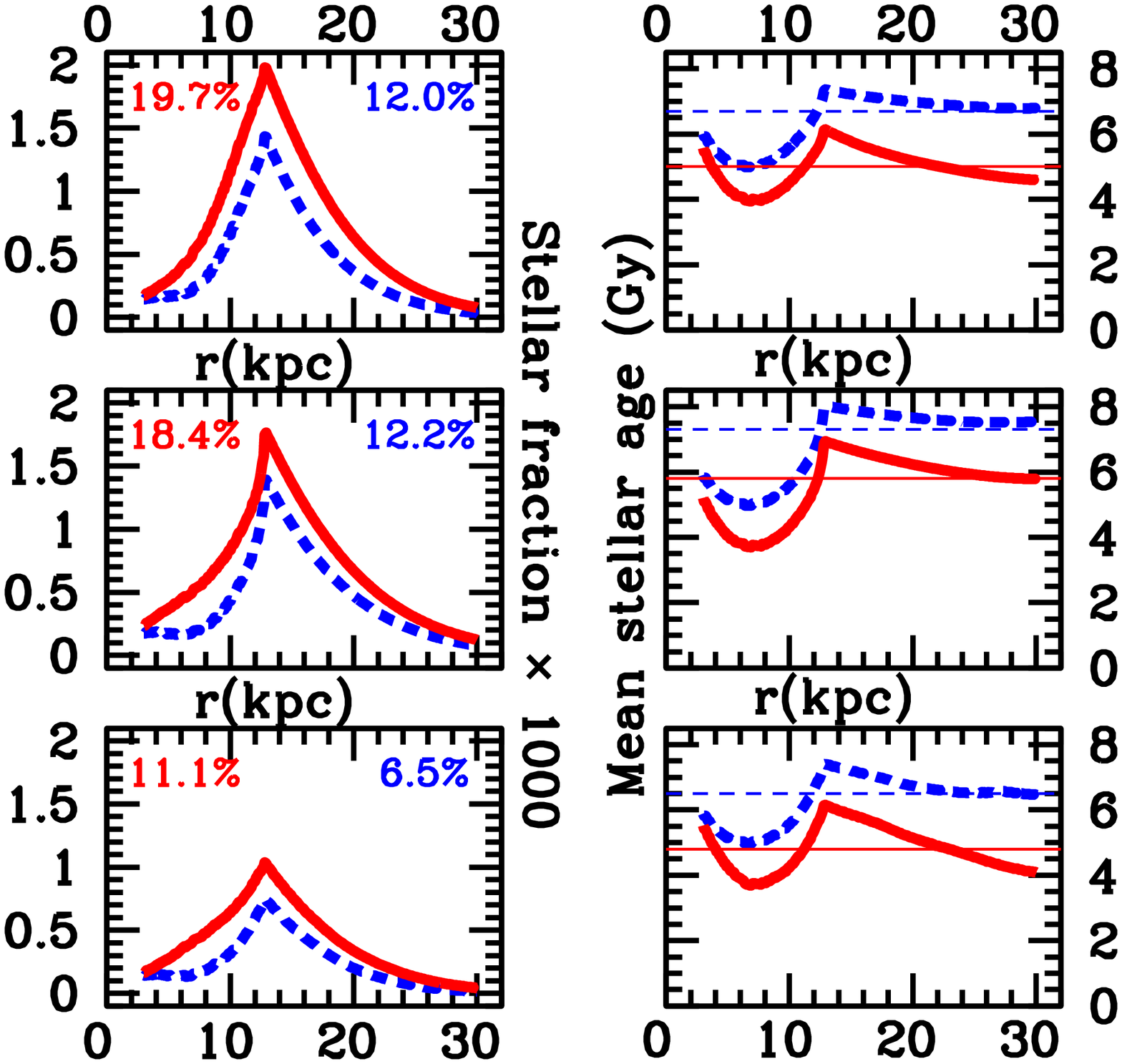} \caption{  Left
panels: $0 - 13$ Gy integrated fraction of stars harboring
terrestrial planets that survive SN explosions and forming according
to three different $Z$ probability (see Figure 5): Lineweaver et al.
(2004, upper panels), Prantzos (2008, middle panels), and extrasolar
planets (2013, lower panels). Lines and percentages as Figure 7.
} 
	\label{fig:zfraction} \end{figure}

	 First, we computed the GHZ of M31 using Lineweaver et al.'s
$Z$ distribution: Earth-like planets, that survive the migration of
Jupiter-like planets, and formed in regions that fulfill the $Z$
dependent  probability shown  in Figure 5 (upper panel). That
probability peaks at log($Z/Z_\odot$) $\sim 0.1$ and is poorly
populated at high metallicities, comparing to the extrasolar planets
probability.

	In Figure 8 (upper panels) we show the M31 GHZ, taking into
account the Lineweaver et al.'s $Z$ distribution, and also we impose
the basic and complex life conditions. The GHZ maximum is located
between 3 and 13 kpc, with ages between 5 and 8 Gy due to that $Z$
probability presents a pronounced maximum between 0.05 and 0.2 dex,
being that probability more important than $P_{STAR}$ at that radii
and times.

	Then, we computed the GHZ adopting the Prantzos' $Z$
distribution to form terrestrial planets, that survive the migration
of gaseous planets: an almost $Z$ independent probability for a wider
$Z$ range, $-1.00 \le $ log($Z/Z_\odot$) $\le +0.50$, (Figure 5,
middle panel).

	In Figure 8 (middle panels)  we show the M31 GHZ assuming 
Prantzos' $Z$ distribution plus the basic and complex life
conditions. The maximum GHZ is located between 3 and 5 kpc, with ages
between 6 and 7 Gy, where and when the $SFR$ is high enough to
form a huge numbers of stars, but not so high to sterilize their
planets. In this case it is possible to find, with medium
probability, younger  (age $ < $4 Gyr) planets located in the inner
disk ($r < 5$ kpc) and older planets (age $> 6$ Gyr) located in the
central disk ($12 < r < 20$ kpc). Since the $P_Z$ is almost constant,
the GHZ behavior is determined mainly by the $SFR$ and SN events.

	 For comparison, we present in Figure 8 (lower panels) the
GHZ adopting the $Z$ distribution shown by the exoplanets. That
probability has a wide peak slightly moved to higher metallicities
(Figure 5, lower panel). The maximum GHZ is located between 3 and 5
kpc, with ages between 6 and 7 Gy, due to that $P_Z$ peaks at
log($Z/Z_\odot$) $\sim 0.2$, where the $P_{STAR}$ is important.

	 In the left panels of Figure 9, for the same GHZs of Figure
8, we plot the $0 - 13$ Gy integrated fraction of stars with
terrestrial planets unsterilized by SN. The numbers at the left and
right corners represent the total percentages of those planetary systems capable of sustaining 
basic life or evolved life, respectively. Both
percentages increase with the mean probability of the $Z$
distribution. The width at half maximum of the stellar fractions are
located at $9.7 < r(kpc) < 18$, $10.0 < r(kpc) < 18$ and $9.0< r(kpc)
< 18$ for Lineweaver et al. (2004), Prantzos (2008), and exoplanets
(2013) distributions, respectively, due to the $P_Z/P_Z^{max}$ values
at high $Z$. For $r < 12.5$ kpc and $r > 12.5$ the decline rate of
the stellar fractions rate reflects the $P_Z$  behavior at super
solar and subsolar metallicities, respectively. Again, the stellar
fractions peak at higher radii than GHZs peak, due to the
time-integration of the $P_{GHZ}(r, t)$ and the area of each
concentric ring.

	In the right panels of Figure 9 we plot the mean age of the
stellar radial distribution present in the left panels. The thin
lines represent the average age of the total stars harboring
Earth-like planets capable of hosting life. For $r > 12$ kpc when: i) $P_Z$ by
Prantzos (2008) is used, the mean stellar ages are higher than those
by Lineweaver et al. (2004) and exoplanets (2013) distributions,
producing average stellar ages higher by 1 Gy, approximately. That
fact is due to that distribution presents a plateau of maximum
probability for low $Z$, increasing the number of metal poor and
older stars. ii) $P_Z$ by exoplanets (2013) distribution is assumed,
the mean ages are lower, because that distribution shows low
probabilities for low $Z$.

	\section{DISCUSSION} \label{sec:discussion}

	\subsection{Chemical Evolution model} \label{sec:discem}

	A chemical evolution model can be computed with three minimum
observational constraints at the present time: total baryonic mass,
gas mass, and the abundance of one chemical element. In this study we
have compiled from the updated literature and well described
$\Sigma_T(r)$ and $\Sigma_{gas}(r)$  distributions, and we have
determined a O/H gradient from {\hii} regions. Based on these quite
restrictive constraints, we obtained a solid chemical evolution model
of the halo and disk of M31.

	The chemical evolution model was built to match the most probable
O/H gradient, [O/H]($r$) = $-$0.015 dex kpc$^{-1}$ $\times$ r(kpc) +
0.44 dex,  determined considering the intrinsic scatter and several
theoretical calibrations (see section 2.1.2). This gradient is caused
by the inside-out scenario: the infall is faster and more abundant in
inner regions, producing a more efficient SFR, which increases the
O/H abundance of the inner parts faster than that of the outer parts.

	In this study, we also explored the gradient determined by
empirical calibrations and proposed by Pilyugin (2001) and Pilyugin
\& Thuan (2005). The gradient obtained through the Pilyugin (2001)
calibration presents the same slope, but its y-intercept value is
lower by 0.42 dex than that obtained from theoretical calibrations.
On the other hand, the gradient obtained by using the Pilyugin \&
Thuan (2005) calibration, [O/H]($r$) =$-$0.008 dex kpc$^{-1}$
$\times$ r(kpc) $-$ 0.23 dex, is 0.007 dex kpc$^{-1}$ flatter and the
y-intercept value lower by 0.67 dex than that obtained from
theoretical calibrations. We were not able to reproduce the Pilyugin
gradient without overestimating the $\Sigma_{gas}(r)$ distribution.

	Peimbert et al. (2007) suggested that the abundances derived from
the R$_{23}$ empirical calibrations could be underestimated due to
the presence of temperature variations in the ionized volume, which
leads to the overestimation of the electron temperatures, and that
the use of empirical calibrations based on optical recombination line
measurements (which are much less affected by temperature variations
and give abundances similar to those obtained using theoretical
calibrations) could reconcile the R$_{23}$ theoretical calibrations
with empirical ones. This is a well known problem in the astrophysics
of photoionized nebulae (see Peimbert et al. 2007 and references
therein for more discussion about this topic).

	The slope of the O/H gradients we computed are in some
way steeper, but are consistent, within the errors, to those derived
from supergiants within 12 kpc of the galactic center reported by
Trundle et al. (2002), which amounts to $-$0.006$\pm$ 0.020 dex
kpc$^{-1}$. Venn et al. (2000) had also found that there was no
oxygen gradient between 10 kpc and 20 kpc; however, their study was
based on abundance determinations for only three A--F supergiants and
their results are therefore inconclusive. Our result is also
consistent with the canonical M31 gradient derived by
\citet{zaritskyetal94} ($-$0.020$\pm$0.007). Very recently, two
exhaustive works on the O/H gradient from {\hii} regions have been
developed: \citet{zuritabresolin12}, from the analysis of 85 {\hii}
regions, computed an O/H gradient of $-$0.023$\pm$0,002 dex
kpc$^{-1}$, which is somewhat steeper than ours. Depending on the
adopted strong line method, they find a central (y-intercept) [O/H]
abundance in the range 0.02$-$0.22, which is 0.22$-$0.42 dex lower than the value we
have adopted. On the other hand, \citet{sandersetal12} have developed
the most extensive work on the O/H gradient in M31 to date; they
computed the O/H gradient using several strong line calibration
methods applied to more than a hundred {\hii} regions, particularly,
applying their prefered strong line method \citep[those
of][]{nagaoetal07} to 192 {\hii} regions, they find an O/H gradient
of $-$0.0195$\pm$0.0055 dex kpc$^{-1}$ with a central
[O/H]=0.44$\pm$0.07, in excellent agreement with the values computed
in this work. However, from direct abundance measurements in 51
planetary nebulae (PNe) they do not find  a significant O/H abundance
gradient. Finally, \citet{kwitteretal12} have recently computed a
radial O/H gradient from PNe data;  they derive  a gradient of
$-$0.011$\pm$0.004 dex kpc$^{-1}$, which is consistent, within the
errors, with our value derived from {\hii} regions.
	We need additionally to bear in mind that, although the values of
the gradient we have obtained are somewhat shallower than typical
values for spiral galaxies, recent results seem to confirm that M31
is a barred spiral galaxy (see Beaton et al. 2007, Athanassoula \&
Beaton 2006 and references therein), which produce shallower
gradients than normal spiral galaxies (Friedli et al. 1994).

	The $\Sigma_{gas}$ obtained from the chemical evolution model
follows a similar pattern to that observed at radii larger than 9
kpc. However, at smaller radii the model predicted more gas than
observed. Since $\Sigma_{gas}$, at a fixed galactocentric distance,
decreases due to the SFR and increases mainly due to infall, we tried
to reproduce at smaller r values the $\Sigma_{gas}$ increasing the
SFR, but the slope of the O/H gradient became steeper than the most
probable one. In addition, we tried to solve the discrepancy between
the model and the observed $\Sigma_{gas}$ at low radii, changing the
radial distribution of the infall, but the agreement with the
observed $\Sigma_{T}$ was lost. It is possible to reduce
$\Sigma_{gas}$  in the inner part by adopting gas inflows to the
galactic bar ( Portinari \& Chiosi 2000,  Spitoni \& Matteucci 2011).
That assumption will be considered in a future article.

	The predicted distribution of the SFR at 13 Gy agrees with the
observed one (see data collected by Yin et al. 2009 and Marcon-Uchida
et al. 2010) for inner regions, but is in disagreement for outer
regions. On the other hand, the agreement between the predicted and
observed $\Sigma_{gas}$ is good for the outer regions, but not for
inner regions. Since the SFR assumed in this study depends on the gas
mass, there is an inconsistency between the observed $\Sigma_{gas}$
and the SFR. Therefore, it is not obvious how to change the two free
parameters of our SFR (see section 2.2) to match both the
$\Sigma_{gas}$ and the SFR distributions with the observed values. In
order to improve the agreement of the obtained $\Sigma_{gas}$ and
SFR, we needed to consider different accretion and star formation
laws based on more complex physical properties for galactic and star
formation.

	Renda et al. (2005), Matsson (2008), Yin et al. (2009), and
Marcon-Uchida et al. (2010) have computed CEMs for the Andromeda
galaxy taking into consideration different assumptions and using
diverse codes, but with a common consideration: two-dimensional
disk model with negligible height, as our model. That is a model
limitation, in particular if we like to study the chemical properties
of the thin and thick disks, which is beyond the scope of this paper.
If a three dimensional model were implemented, the general properties
of the disk would not be so different, because the inside-out nature
of disc growth (main assumption in our models) is obtained as a
consequence from cosmological hydrodynamical simulations (Brook et
al. 2012).

	These models were built to reproduce different O/H gradients, 
which show higher dispersion or error, with O/H differences at a
given $r$ value in the 0.4 to 0.7 dex range; consequently, these
models are not constrained as well as our model. In particular the
agreements of the Marcon-Uchida et al. model are poorer or marginal.

	We also made a comparative study of the chemical evolution models
and the observational constraints of the Milky Way galaxy with the
Andromeda galaxy, and we noticed that:

	\begin{enumerate} \item The maximum average of the Fe/H
abundances present in the halo stars of M31 corresponds to
log($Z/Z_\odot$)= $-$0.5, while in the MW it amounts to $-$1.6 (Koch
et al. 2008). In order to reproduce the higher $Z$ value in the halo
stars of M31, our model requires, in the halo phase, a more efficient
infall ($\tau_h$ is about 5 times lower) and SFR (two orders of
magnitudes higher) than those for the MW model computed with similar
assumptions (Carigi et al. 2005, Carigi \& Peimbert 2011). \item The
adopted gradient for M31, [O/H]($r$) = $-$0.015 dex kpc$^{-1}$
$\times$ r (kpc) + 0.44 dex, is 0.03 dex flatter and 0.06 dex higher
in the central value than the O/H gradient of the Milky Way disk
determined by Esteban et al. (2005) from optical oxygen recombination
lines. In order to reproduce the flatter  M31 gradient, our model
needed, in the disk phase, a less marked inside-out formation, i.e.
slower infall in the inner regions and faster accretion in the outer
parts compared to the infall considered in CEMs of the MW. Moreover,
to reach the higher O/H values, the model required a more active star
formation history than that assumed for the MW disk (Carigi \&
Peimbert 2011), resulting in a lower  $\Sigma_{gas}(r)$ in M31 than
in the MW, in agreement with the observed distribution in both
galaxies. \end{enumerate}

	\subsection{Galactic Habitable Zone} \label{sec:disghz}

	The chemical evolution model discussed above provides the
evolution and location of the heavy elements of which planets are
formed, the stellar surface density, and the rate of neighboring
SN that may sterilize life on the planets. The age and the location
of potential life-bearing planets inside a galaxy depend on the $Z$
range assumed to form those planets and on the number of stars
harboring planets.

	Using the $Z$ distribution shown by the exoplanet encyclopaedia (2013) we studied
the SN effects on the GHZ per $pc^2$ assuming three different
permanent planet sterilization conditions: $RSN(r,t)$,
$<RSN(r,t)>_{planet \ age}$, and  $1/2 <RSN(r,t)>_{4.5 Gy}$ is higher
than $<RSN_{SV}>$ From Figures 6 and 7, we note that SN effects are
more important in the inner disk during the first half of the
galactic evolution. Moreover, we figure out when life's resistance to
SN danger increases: i) the location of the most probable GHZ per
$pc^2$ shifts from the central concentric rings ($12 < r(kpc) < 18$)
to the inner rings ($3 < r(kpc) < 6$), ii) the average stellar age
increases, and iii) the total number of star harboring unsterilized
planets increases.

	The SN condition for extinguishing the life on a planet is the
most uncertain restriction, because it is difficult to estimate the
resistance of life to SN explosions, which depends on unknown
biogenic and astrophysical factors, such as the minimum SN-planet
distance that allows life to survive, the type of atmosphere and
oceans where life could evolve, and the biological properties of the
different types of life, among others factors. Due to life on
the Earth has been able to recover massive extinctions, caused by
different events, including neighbor SN events, we consider that the
double of the average SN rate of the solar neighborhood, since the
Earth formed ($<RSN(r,t)>_{4.5 Gy}$) is the maximum average that life
can withstand on an planet.

	Assuming that life disappear forever if  $<RSN(r,t)>_{4.5 Gy}$ is
higher than $2 \times <RSN_{SV}>$, we studied the Z effects on the
GHZ per $pc^2$ considering three different  $Z$ probability
distribution to form Earth-like planets: Lineweaver et al. (2004),
Prantzos (2008), and extrasolar planets (2013). From Figures 8 and 9,
we note that Z effects are more important in the inner disk during
the second half of the galactic evolution. Moreover, we figure out
that when: i)  the distribution peaks at higher $Z$, the location of
the most probable GHZ per $pc^2$ shifts from the inner disk  ($3 <
r(kpc) < 12$) to the very inner rings ($3 < r(kpc) < 5$), ii) the $Z$
distribution is weighted to low $Z$, the average stellar age
increases, and iii) the half maximum value of the distribution
increases, the total number of star harboring terrestrial planets
increases.
	It is difficult to quantify the role of metallicity in planet
formation. At present, theoretical models fail to explain in detail
the physical processes in protoplanetary disks and in particular the
metallicity effect on the planet mass (rocky like Earth or gaseous
like Jupiter). Furthermore, the observational data of detected
extrasolar planets do not show a clear trend between the stellar
metallicity and the planet mass that orbit around it. Recent
Jenkins et al. (2013) have increased the number of low-mass planets
around metal-rich stars. However, Adibekyan et al. (2012) confirm an
overabundance in giant-planet host stars with high $Z$. 
	Also, it is noticeable that in many detected extrasolar planetary
systems, Earth-like and Jupiter-like planets coexist (The Extrasolar
Planets Encyclopedia). Moreover, Raymond et al. (2006) have simulated
terrestrial planet growth during and after giant planet migration.
Observational studies on the circumstellar disk  in young stellar
clusters of low and solar metallicities suggest that the disk
lifetime shortens with decreasing $Z$, thereby reducing the time
available for planet formation (Yasui et al. 2010). Based on
circumstellar disk models and dust grain properties, Johnson \& Li
(2012) conclude that the first Earth-like planets likely formed from
gas with metallicities $Z > 0.1 \ Z_\odot$.

	For these reasons we preferred the $Z$ distribution shown by
exoplanet-harboring stars as the metallicity condition for the GHZ
instead of undertaking a probability study of planet formation and
migration as a function of metallicity, as did Lineweaver (2001),
Prantzos (2008), and Gowanlock et al. (2011).

	Based on the previous discussion, we deduce that: the most
probable GHZ per $pc^2$ is located in the galactic disk of M31
between 3 and 7 kpc and ages between 6 and 7 Gy, assuming both basic
and complex life (see Figures 6 and 8, lower panel), but the
highest amount of stars harboring planets that survive SN is in an 
annular region of 13 kpc radius and 2 kpc width, with mean ages of
5.8 and 7.0 Gy, for basic and complex life, respectively.

	In GHZ studies the galactic formation scenario is crucial for
finding the location and number of planetary systems with at least a
terrestrial planet, because the galactic formation scenario
determines the infall rate, that provides the gas to form stars, when
those stars die enrich the ISM with heavy chemical elements, from
that metals planets form, and those planets far enough from
SN, survive.

	\subsubsection{ Comparison with GHZ of the MW}
\label{sec:disghz}

	 The GHZs for the MW were obtained by Lineweaver et al.
(2004), Prantzos (2008) and Gowanlock et al. (2011) from similar
chemical evolution models,  built on the inside-out scenario,
consequently, the SFR(r,t) and Z(r,t) can be considered equal in all
MW GHZs. Therefore the differences among those GHZs are caused by 
the differences in the assumed probabilities, mainly in the $Z$
distributions to form terrestrial planets. Since one of the scopes
of this paper is to find  the differences in the habitability of
Milky Way and Andromeda galaxies, we computed in section 3.3.2 GHZs
for M31, based on the $Z$ probability used by Lineweaver et al. and
Prantzos (Gowanlock et al's distribution is almost constant, as
Prantzos' one).

	First, Lineweaver et al. (2004) found that the GHZ of the MW is
located in the inner-central Galactic disk, like our GHZ of M31,
based on the same $Z$ probability (see Figure 8, upper panel). Their
GHZ with probabilities higher than 68\%, and complex or basic life,
is a wide ring between $r=4$ kpc and 11 kpc, which area corresponds
to 26\% of the Galactic disk area considered by those authors ($\sim
\pi (20^2 - 2.5^2) kpc^2$). In our work, the GHZ with $P_{GHZ} >
0.68$ is a ring between 3 and 12 kpc that corresponds to 15\% of the
galactic disk of M31 ($\sim \pi (30^2 - 3^2) kpc^2$). According to
Lineweaver et al. less than 10\% of stars formed in MW have a
probability higher than 0.68 to harbor terrestrial planet capable of sustaining
complex life, but in our GHZ that percentage is 1.6\% and
corresponds, approximately, to the stellar fraction between 3 and 10
kpc (see Figure 9, upper panel).

	Second, Prantzos (2008) found that the GHZ, only for basic life,
is located in the inner Galactic disk, like our GHZ of the M31, based
on the same $Z$ probability (see Figure 8, middle panel). His GHZ,
with probabilities higher than 50\% is a ring between $r=2$ kpc and 8
kpc, that corresponds to 17 \% of the Galactic disk area considered
by him ($\sim \pi (19^2 - 2^2) kpc^2$). In our work, the GHZ with
$P_{GHZ} > 0.50$ is a ring between 3 and 12 kpc that corresponds to 15\%
of the M31 galactic disk. Unfortunately, he does not mention the
percentage of stars suitable for life, but in our GHZ that percentage
is 1.7 \% and corresponds to stellar fractions between 0.2 and 0.3
values in the 3-12 kpc range.

	Then, Gowanlock et al. (2011) found that the GHZ, only for
complex life, is located toward the inner Galaxy. It is not easy to
compare their GHZ and ours, because they did more precise and
complete computations, assuming, e.g. three dimensional disk,
probabilities from SN survival condition based on the planet-SN
distance, SNII and SNIa, and tidal locking. If we focus only on the
common ingredients ($P_{STAR}$ and $P_Z$), we confirm that their
$P_{STAR}$ is similar to ours, due to inside-out scenario, but their
$P_Z$ is similar to Prantzos one, because their probability is almost
independent on $Z$. Therefore, if a comparison could be done is with
our GHZ that considers the $Z$ probability by Prantzos (2008, see
Figure 8, middle panel). The width of the half maximum probability of
their GHZ is located between 3 and 8 kpc, that corresponds to 25 \%
of the Galactic disk area considered by them ($\sim \pi (15^2 - 3^2)
kpc^2$), but our percentage is 15 \%. According to Gowanlock et al.
1.2 \% of all stars formed in MW host a planet suitable for life, but
based on Figure 9 (middle panel) our percentage is a factor of 10
larger (12.2 \%).

	Unfortunately, the comparison of the GHZs  between the MW and
M31, based on stellar fractions and galactic rings where stars
harboring habitable planets are located, was not conclusive. The only
property that we could infer is the position of the most likely GHZ,
which depends on the $Z$ probability assumed. A more successful
comparative study will be obtained when the GHZ of the MW is computed
from identical constraints. That comparison will be shown in a future
paper and will allow to explore the effect on the GHZ of the
intrinsic differences among spiral galaxies.

	Throughout this section we have emphasized that the chemical
content is one of the most restrictive condition to the GHZ.
Furthermore, we have mentioned that the slope of the $Z$ gradient is
determined by the efficiency of inside-out scenario, whereas the
y-intercept values of the gradient depend on by the amount of
accreted gas and the star formation efficiency. Therefore, we stress
that the manner in which a galaxy is assembled and its stars formed
determine the size and age of the GHZ.

	\section{Conclusions} \label{sec:thend}

	Based on O/H values of {\hii} regions, we obtained current O/H
gradients with different empirical and theoretical calibrations. In
the presence of intrinsic scatter, we computed the most probable
gradient for M31 from theoretical calibrations based on the R$_{23}$
method and we found that [O/H]($r$) = $- 0.015 \pm 0.003 $dex
kpc$^{-1} \times r(kpc) + 0.44 \pm 0.04$ dex. The slope of the
gradient is 0.03 dex kpc$^{-1}$ flatter, and the value at the center
of the galaxy is 0.06 dex higher, than those values of the O/H
gradient of the Milky Way disk.

	The chemical evolution model built to reproduce our O/H gradient
of the galactic disk matches the current radial distribution of the
gas mass, $\Sigma_{gas}(r)$, for the outer regions quite well, but it
fails for the inner ones. On the other hand, the current radial
distribution of the star formation rate, SFR($r$), fits for the inner
parts, but not for the outer ones. Therefore, in order to improve the
agreements, the model would require more complex galactic and star
formation histories.

	The model cannot reproduce the O/H gradient computed by empirical
methods, which is 0.42 dex lower for the central value than the
gradient obtained by theoretical methods, unless the other
observational constraints are considerably modified.

	Based on the chemical evolution model we obtained, for the
first time, the Galactic Habitable Zone per $pc^2$ (GHZ) of M31,
considering three space requirements (per surface unit):  i)
sufficient metallicity for planet formation with a probability law
that follows the $Z$ distribution shown by exoplanets, ii) high
number of stars that may be potential home for life,  and iii) an
average SN rate similar to that permitting the existence of life on
the Earth; and two time requirements: the existence of basic life
(like cyanobacteria), and the development of evolved life (like
humans).

	 The GHZ of M31 with high probability is located between 3
and 7 kpc on planets with ages between 6 and 7 Gy, approximately.
Assuming the area of each ring, the maximum number of stars, of all
ages and harboring Earth-like planets, is in a ring located between
12 and 14 kpc with mean ages of $\sim$ 6 Gy and $\sim$ 7 Gy for
planet capable of sustaining basic and complex life, respectively.

	The GHZ of M31 with high and medium probability is located
between 3 and 14 kpc (the inner half of the disk) on planets with
ages between 3 and 9 Gy, approximately. The width at half maximum of
number of stars, of any ages and harboring Earth-like planets suitable for
basic life, is in an annular region located between 9 and 18 kpc,
with mean age of $\sim$ 5 Gy. That annular region corresponds to 27
\% of the galactic disk area ($\sim \pi (30^2 - 3^2) kpc^2$). On the
other hand, that width for complex life is located between 10 and 18
kpc, which corresponding to  25 \% of the galactic disk, and the mean
age of  that annular region is $\sim$ 6.5 Gy. According our
computation 11 \% of the stars formed in M31 may have planets capable of hosting
basic life, and 6.5 \% complex life.

	SN effects are important only in the inner half disk and during
the first half of the galactic evolution, but $Z$ effects are more
important during the second half of the galactic evolution.

	In GHZ studies, the $Z$ restriction is crucial for finding the
location of Earth-like planets, but the SN survival condition allows
us to compute the location of those planets that survive SN events.

	 Based on the previous GHZs of the MW and the present GHZ of
the M31, we are not able to compare the GHZs of those spiral
galaxies, due to the GHZs were computed with different  constraints.

	\vskip 1cm

	Part of this work was submitted by Sof\'{\i}a Meneses Goytia to
the Master's Programme in Chemical Sciences at the Universidad
Nacional Aut\'onoma de M\'exico.

	\section{Acknowledgments} \label{sec:thanks}

 The authors wish to thank the anonymous referee for a
careful review of the manuscript and for helpful suggestions, which
improved the quality of this paper significantly.
	The authors thank: C. Esteban for his timely suggestion
concerning the O/H gradient, M. Peimbert and A. Segura for a critical
reading of the manuscript, and T. Mahoney for revising the english
text. L. Carigi thanks E. M. Berkhuijsen for kindly providing
information about gas data of M31 and detailed explanations of how to
compute the hydrogen gas mass. S. Meneses-Goytia thanks T.J.L. de
Boer and S. C. Trager for their careful and detailed review of the
present paper. J. Garc\'{\i}a-Rojas acknowledges partial support from
the project AYA2007-63030 and AYA2011-22614 of the Spanish Ministerio de Educaci\'on y
Ciencia and from a UNAM postdoctoral grant. L. Carigi thanks the
funding provided by the Ministry of Science and Innovation of the
Kingdom of Spain (grants AYA2010-16717 and AYA2011-22614). This work
was partly supported by grant 129753 from CONACyT.


	\end{document}